%% file: main.tex
\documentclass[10pt,aps,pra,twocolumn,superscriptaddress,floatfix,nofootinbib]{revtex4-2}

\makeatletter
 \let\@join\undefined %
\makeatother

\usepackage{amsmath, amsfonts, amssymb}
\usepackage{bm, bbm, braket, accents}
\usepackage{graphicx}   
\usepackage[usenames,dvipsnames]{xcolor}
\usepackage{lipsum} 
\usepackage{tikz,tikz-3dplot}
\usepackage{rubikcube}
\usetikzlibrary{quantikz}
\usepackage{pgfplots}
\pgfplotsset{compat=newest}
\usepackage{mathtools}

\usepackage{transparent}
\usepackage{standalone}
\usepackage{enumitem}
\usepackage{soul}

\makeatletter
\def\@bibdataout@aps{%
 \immediate\write\@bibdataout{%
  @CONTROL{%
   apsrev41Control,author="08",editor="1",pages="0",title="0",year="1",eprint="1"%
  }%
 }%
 \if@filesw
  \immediate\write\@auxout{\string\citation{apsrev41Control}}%
 \fi
}%
\makeatother 

\newcommand{\phm}{\phantom{-}}

\newcommand\SWAP{\mathrm{SWAP}}
\newcommand{\ip}[2]{\langle{#1}|{#2}\rangle}
\newcommand{\op}[2]{\ket{#1}\!\bra{#2}}

\newcommand{\nbym}[2]{${#1\times #2}$}

\newcommand{\halfswap}{square root SWAP}

\newcommand{\HS}{\mathbb{H}}

\usepackage[breaklinks=true]{hyperref}
\hypersetup{
  colorlinks   = true, 
  urlcolor     = blue, 
  linkcolor    = blue, 
  citecolor    = red 
}
\usepackage[normalem]{ulem}
\usepackage[capitalise]{cleveref} 

\definecolor{yc}{HTML}{EFF11F}
\definecolor{gc}{HTML}{8BFE61}
\definecolor{gcd}{HTML}{3ee301} 
\definecolor{bc}{HTML}{50E4EF}
\definecolor{bcd}{HTML}{13C0CD} 
\definecolor{oc}{HTML}{FC640A}
\definecolor{rc}{HTML}{FB3042}
\definecolor{wc}{HTML}{ffffff}
\definecolor{kc}{HTML}{c6c6c6} 

\newcommand{\g}[1]{{\color{gcd} g_{#1}}}
\newcommand{\blu}[1]{{\color{bcd}  b_{#1}}}

\newcommand{\cell}[8]{%
\begin{tikzpicture}
 \node[rectangle,fill=#5, minimum width =10pt, minimum height =10pt ] at (-5pt,5pt)  {\tiny #1};
 \node[rectangle,fill=#6, minimum width =10pt, minimum height =10pt ] at (5pt,5pt)   {\tiny #2};
 \node[rectangle,fill=#7, minimum width =10pt, minimum height =10pt ] at (-5pt,-5pt) {\tiny #3};
 \node[rectangle,fill=#8, minimum width =10pt, minimum height =10pt ] at (5pt,-5pt)  {\tiny #4};
 \draw[step=10pt,black,thin] (-10pt,-10pt) grid (10pt,10pt);
\end{tikzpicture}
}

\newcommand{\solved}{\big|\raisebox{-0.2em}{\resizebox{11pt}{!}{\cell{}{}{}{}{gc}{gc}{bc}{bc}} }\!\!\big\rangle}

\newcommand{\zero}[1]{\left |\raisebox{-0.42em}{\resizebox{#1}{!}{\cell{}{}{}{}{gc}{gc}{bc}{bc}}} \! \right \rangle \!}
\newcommand{\zerod}[1]{\left \langle\raisebox{-0.42em}{\resizebox{#1}{!}{\cell{}{}{}{}{gc}{gc}{bc}{bc}}} \! \right | }
\newcommand{\one}[1]{\left |\raisebox{-0.42em}{\resizebox{#1}{!}{\cell{}{}{}{}{bc}{bc}{gc}{gc}}} \! \right \rangle \!}
\newcommand{\oned}[1]{\left \langle\raisebox{-0.42em}{\resizebox{#1}{!}{\cell{}{}{}{}{bc}{bc}{gc}{gc}}} \! \right | }
\newcommand{\two}[1]{\left |\raisebox{-0.42em}{\resizebox{#1}{!}{\cell{}{}{}{}{bc}{gc}{bc}{gc}}} \! \right \rangle \!}

\newcommand{\three}[1]{\left |\raisebox{-0.42em}{\resizebox{#1}{!}{\cell{}{}{}{}{gc}{bc}{gc}{bc}}} \!\right \rangle \!}

\newcommand{\four}[1]{\left |\raisebox{-0.42em}{\resizebox{#1}{!}{\cell{}{}{}{}{bc}{gc}{gc}{bc}}} \! \right \rangle\!}

\newcommand{\five}[1]{\left |\raisebox{-0.42em}{\resizebox{#1}{!}{\cell{}{}{}{}{gc}{bc}{bc}{gc}}} \! \right \rangle\! }
\newcommand{\fived}[1]{\left \langle\raisebox{-0.42em}{\resizebox{#1}{!}{\cell{}{}{}{}{gc}{bc}{bc}{gc}}} \! \right |}

\newcommand{\puzzIPone}[1]{\left \langle\raisebox{-0.42em}{\resizebox{#1}{!}{\cell{}{}{}{}{bc}{gc}{gc}{bc}}} \! \right | 
\!\left.\raisebox{-0.42em}{\resizebox{#1}{!}{\cell{}{}{}{}{gc}{gc}{bc}{bc}}} \! \right \rangle \!}

\newcommand{\puzzIPtwo}[1]{\left \langle\raisebox{-0.42em}{\resizebox{#1}{!}{\cell{}{}{}{}{gc}{gc}{bc}{bc}}} \! \right | 
\!\left.\raisebox{-0.42em}{\resizebox{#1}{!}{\cell{}{}{}{}{bc}{gc}{gc}{bc}}} \! \right \rangle \!}

\crefformat{equation}{Eq.~(#2#1#3)} 
\crefformat{section}{Sec.~#2#1#3} 
\Crefformat{equation}{Equation~(#2#1#3)}
\crefformat{figure}{Fig.~#2#1#3}
\crefrangeformat{equation}{Eqs.~#3(#1)#4--#5(#2)#6}
\Crefformat{section}{Section~#2#1#3}

\begin{document}

\title{Quantum permutation puzzles with indistinguishable particles}

\author{Noah Lordi}
\thanks{These two authors contributed equally.}
\affiliation{Department of Physics, University of Colorado Boulder, 2000 Colorado Ave, Boulder, CO 80309}
\author{Maed\'ee Trank-Greene}
\thanks{These two authors contributed equally.}
\affiliation{Department of Applied Mathematics, University of Colorado Boulder, Boulder, CO 80309}
\author{Akira Kyle}
\affiliation{Department of Physics, University of Colorado Boulder, 2000 Colorado Ave, Boulder, CO 80309}
\author{Joshua Combes}
\affiliation{Electrical, Computer, and Energy Engineering, University of Colorado Boulder, Boulder, CO 80309}
\affiliation{Schools of Physics and Mathematics, University of Melbourne, VIC 3010, Australia}

\date{\today}

\begin{abstract}
Permutation puzzles, such as the Rubik's cube and the 15 puzzle, are enjoyed by the general public and mathematicians alike. 
Here we introduce quantum versions of permutation puzzles where the pieces of the puzzles are replaced with indistinguishable quantum particles. The moves in the puzzle are achieved by swapping or permuting the particles. 
We show that simply permuting the particles can be mapped to a classical permutation puzzle, even though the identical particles are entangled. However, we obtain a genuine quantum puzzle by adding a quantum move---the square root of SWAP. The resulting puzzle cannot be mapped to a classical permutation puzzle.  We focus predominately on the quantization of the \nbym{2}{2} slide puzzle and briefly treat the $2\times 2 \times 1$ Rubik's cube.
\end{abstract}

\maketitle

\section{Introduction}\label{sec:intro}

Permutation puzzles~\cite{mulholland2016permutation}, such as the Rubik's cube and the 15-puzzle, involve scrambling the pieces, with the objective of returning the puzzle to its original arrangement. All of these puzzles are solved by permuting the pieces in groups determined by the connectivity of the puzzle. For the Rubik's cube, the basic permutation is a rotation about a face. For the 15-puzzle---an example of a ``slide puzzle''---the basic permutation is swapping or sliding a piece into the location of a hole; see \cref{fig:clas_perm_puzz}. Our goal is to develop quantum versions of these puzzles.

Previous quantum computing research has linked the Rubik's cube to quantum walks for encryption~\cite{ZhaoTian2022}, {analogies of} scrambling in many-body dynamics~\cite{ThomsonEisert2024}, Clifford gate synthesis~\cite{BaoHartnett2024}, energy-level diagrams~\cite{wang2024}, and nonlocal entangled states for quantum secret sharing~\cite{ShiHu2020}. Additionally,~\citet{Corli_2023} used reinforcement learning to solve the classical cube by embedding its group into a boson-fermion model.
Similarly, the 15 puzzle has received attention due to its relationship to itinerant ferromagnetism~\cite{BobrowStubisLi2018}. These analogies are useful for understanding and explaining deep results. 
However, our goal is to create a family of quantum puzzles whose primary purpose is quantized entertainment.

In this article, we construct quantum puzzles inspired by slide puzzles and the Rubik's cube. To make a quantum puzzle, we will replace the pieces with quantum particles. A single species of identical particles~\cite{Pauli1940,Spivak2022} will replace all the blue pieces and so on for all the other colors. All particles of the same ``color" will be indistinguishable from one another, but all particles of different colors will be distinguishable. As the pieces are permuted, we must correctly account for the exchange statistics of the identical particles.

\begin{figure}[!th]
\flushleft
{\footnotesize (a) Rubik's cube}\\ \vspace{2pt}
\begin{minipage}{.45\columnwidth}
        \centering
        \includestandalone[width=0.75\columnwidth]{cube_solved}\\\vspace{-2pt}
        {\footnotesize Solved}
\end{minipage}%
\begin{minipage}{.45\columnwidth}
        \centering
        \includestandalone[width=0.75\columnwidth]{cube_scram}\\\vspace{-2pt}
        {\footnotesize Scrambled}
\end{minipage}\\
\flushleft
{\footnotesize (b) Slide puzzle}\\ \vspace{2pt}
\begin{minipage}{.45\columnwidth}
        \centering
        \includestandalone[width=0.45\columnwidth]{slide_solved}\\\vspace{-2pt}
        {\footnotesize Solved}
\end{minipage}%
\begin{minipage}{.45\columnwidth}
        \centering
        \includestandalone[width=0.45\columnwidth]{slide_scram}\\\vspace{-2pt}
        {\footnotesize Scrambled}
\end{minipage}
\flushleft
{\footnotesize (c) Quantum puzzle}\\\vspace{2pt}
\begin{minipage}{.25\columnwidth}
        \centering
        \includestandalone[width=0.6\columnwidth]{slide_quant}\\\vspace{-2pt}
        {\footnotesize Solved}
\end{minipage}%
\begin{minipage}{.25\columnwidth}
        {$ \xRightarrow[\text{scramble}]{\text{quantum}}$}
\end{minipage}%
\begin{minipage}{.5\columnwidth}
        \centering
        \raisebox{-5pt}{\includestandalone[width=0.8\columnwidth]{slide_super}} \\ \vspace{2pt}
        {\footnotesize Superpostion}
\end{minipage}\\
    \caption{Permutation puzzles. In (a) is the familiar $3\times 3 \times 3$ Rubik's cube. In (b) is a slide puzzle (like the 15 puzzle), here the gray square indicates a missing puzzle piece. The other pieces can be moved by sliding them in to the empty space. In (c) we give a cartoon of our quantum puzzle. We propose replacing the colored faces with a quantum particle. If the faces have the same color the corresponding particles must be indistinguishable. Then by performing fractional exchanges of identical particles we can build up non-classical puzzle states.
    }\label{fig:clas_perm_puzz}
\end{figure}

We focus on a 2D version of a permutation puzzle with identical particles, which is related to the 15 puzzle. We show in \cref{sec:perm_puzz} that replacing pieces with quantum particles results in a new effectively classical puzzle that is finite-dimensional. Fortunately, the quantum structure of the puzzle permits new actions that are not possible classically. In \cref{sec:quant_perm_puzz} we add a new move, a square root SWAP, which allows the state of the puzzle to leave the subspace of classically allowed states. Then in \cref{sec:solns} we investigate solving strategies and compare how a solver with access to square root SWAPs can outperform solvers with only classical actions. We then consider several extensions to the puzzle. First, in \cref{sec:universality} we explore adding more moves to allow our puzzle to, in principle, reach all states in the puzzle Hilbert space. Second, in \cref{sec:board_size} we consider larger puzzles with more particles.  Third, in \cref{sec:cube} we demonstrate how to generalize our 2D puzzle to 3D with a specific example of the $2\times 2\times 1$ Cube puzzle. Finally, we conclude with some open questions in \cref{sec:concl}.

\section{Permutation Puzzles}\label{sec:perm_puzz}

Consider the following classical \nbym{2}{2} slide--like puzzle with two blue tiles and two green tiles
 \begin{equation}
\cell{1}{2}{3}{4}{gc}{gc}{bc}{bc} \,.
\end{equation}
Here, the colored tiles are labeled 1,2,3,4 so they are distinguishable.
The most basic operation in permutation puzzles is some method of permuting the pieces. For the slide puzzle, these permutations take the form of sliding a piece into a hole. The first departure of our puzzle from a regular slide puzzle is that we do not require a missing piece, instead we allow the swapping of pieces. In particular, we allow the following SWAP moves
\begin{equation}\label{eq:swaps}
\begin{aligned}
\cell{1}{2}{3}{4}{gc}{gc}{bc}{bc} \, \raisebox{7pt}{$\xrightarrow{\,\text{Up}\,\,}$} &\,\,\cell{2}{1}{3}{4}{gc}{gc}{bc}{bc}\,,\quad \cell{1}{2}{3}{4}{gc}{gc}{bc}{bc} \,\raisebox{7pt}{$\xrightarrow{\text{Down}}$}\,\,\cell{1}{2}{4}{3}{gc}{gc}{bc}{bc}\, ,\\
\cell{1}{2}{3}{4}{gc}{gc}{bc}{bc} \, \raisebox{7pt}{$\xrightarrow{\text{Left}\,} $}&\,\,\cell{3}{2}{1}{4}{bc}{gc}{gc}{bc}\,,\quad \cell{1}{2}{3}{4}{gc}{gc}{bc}{bc} \, \raisebox{7pt}{$\xrightarrow{\text{Right}}$}\, \,\cell{1}{4}{3}{2}{gc}{bc}{bc}{gc}\,;
\end{aligned}
\end{equation}
 however we do not allow SWAPs between diagonal elements. Given these actions, it is easy to compute that there are $4!=24$ arrangements of the board. 
 
 Next, we remove the labels to make a stronger analogy with identical particles. However, we insist that the puzzle respects orientation in the same way that most slide puzzles do. This means that rotating the board is not an allowed action, e.g. a $90^{\circ}$ rotation about the center
\begin{equation}
    \begin{tikzpicture}
     \node[rectangle,fill=gc, minimum width =10pt, minimum height =10pt ] at (-5pt,5pt)  {};
    \node[rectangle,fill=gc, minimum width =10pt, minimum height =10pt ] at (5pt,5pt)   {};
    \node[rectangle,fill=bc, minimum width =10pt, minimum height =10pt ] at (-5pt,-5pt) {};
    \node[rectangle,fill=bc, minimum width =10pt, minimum height =10pt ] at (5pt,-5pt)  {};
    \draw[step=10pt,black,thin] (-10pt,-10pt) grid (10pt,10pt);
    \draw[thick, ->] (16pt,2pt) arc (0:80:0.5cm);
    \end{tikzpicture}
    \,\,\,\raisebox{0.75em}{$\xrightarrow{\text{rotation}}$}\,\,\,\, \cell{}{}{}{}{gc}{bc}{gc}{bc} \quad \raisebox{6pt}{(\text{not allowed})\,.}
\end{equation}
When this is taken into account and when the labels are removed there are now  $4!/(2!)^2=6$ unique board states:
\begin{equation} \label{eqn:classicalStates}
\raisebox{-0.42em}{\cell{}{}{}{}{gc}{gc}{bc}{bc}}, \,\,
 \raisebox{-0.42em}{\cell{}{}{}{}{bc}{bc}{gc}{gc}} , \,\,
\raisebox{-0.42em}{\cell{}{}{}{}{bc}{gc}{bc}{gc}}, \,\,
\raisebox{-0.42em}{\cell{}{}{}{}{gc}{bc}{gc}{bc}} , \,\,
\raisebox{-0.42em}{\cell{}{}{}{}{bc}{gc}{gc}{bc}} , \,\,
\raisebox{-0.42em}{\cell{}{}{}{}{gc}{bc}{bc}{gc}}  \, .
\end{equation}
It is also clear that using the actions in \cref{eq:swaps}, any board state can be generated from any other board state.

With the classical board understood we will replace the classical tiles with identical quantum particles. These particles could be sets of identical fermions or bosons. We are interested in the possible unique states of four particles on the board. Two of the particles are green and two are blue. The blue and green particles are distinguishable from each other. But particles with the same color are indistinguishable.

We denote the solved state of the quantum puzzle by
\begin{equation}
  \zero{15pt}.
\end{equation}
Because some of the particles are identical we must symmetrize or anti-symmetrize the particles, for this reason we momentarily re-introduce particle labels.  For identical fermions or bosons the appropriately symmetrized solved state, up to normalization, is
\begin{subequations}\label{eq:board0symm}
\begin{alignat}{5}
 \left |\raisebox{-0.42em}{\resizebox{15pt}{!}{\cell{\phantom{1}}{\phantom{1}}{\phantom{1}}{\phantom{1}}{gc}{gc}{bc}{bc}}} \!\! \right \rangle &{}\propto{}&
       \left |\raisebox{-0.42em}{\resizebox{15pt}{!}{\cell{1}{2}{1}{2}{gc}{gc}{bc}{bc}}} \!\! \right \rangle  
&{}-{}&\left |\raisebox{-0.42em}{\resizebox{15pt}{!}{\cell{2}{1}{1}{2}{gc}{gc}{bc}{bc}}} \! \right \rangle 
&{}-{}&\left |\raisebox{-0.42em}{\resizebox{15pt}{!}{\cell{1}{2}{2}{1}{gc}{gc}{bc}{bc}}} \!\! \right \rangle  
&{}+{}&\left |\raisebox{-0.42em}{\resizebox{15pt}{!}{\cell{2}{1}{2}{1}{gc}{gc}{bc}{bc}}} \! \right \rangle \,\, (\text{fermionic}), \label{eq:board0symm_ferm}\\
 \left |\raisebox{-0.42em}{\resizebox{15pt}{!}{\cell{\phantom{1}}{\phantom{1}}{\phantom{1}}{\phantom{1}}{gc}{gc}{bc}{bc}}} \!\! \right \rangle &{}\propto{}&
       \left |\raisebox{-0.42em}{\resizebox{15pt}{!}{\cell{1}{2}{1}{2}{gc}{gc}{bc}{bc}}} \!\! \right \rangle  
&{}+{}&\left |\raisebox{-0.42em}{\resizebox{15pt}{!}{\cell{2}{1}{1}{2}{gc}{gc}{bc}{bc}}} \! \right \rangle 
&{}+{}&\left |\raisebox{-0.42em}{\resizebox{15pt}{!}{\cell{1}{2}{2}{1}{gc}{gc}{bc}{bc}}} \!\! \right \rangle  
&{}+{}&\left |\raisebox{-0.42em}{\resizebox{15pt}{!}{\cell{2}{1}{2}{1}{gc}{gc}{bc}{bc}}} \! \right \rangle  \,\, (\text{bosonic}). \,\,\,\, \label{eq:board0symm_bose}
\end{alignat}
\end{subequations}
In representing the identical particles this way we have assumed that the board is represented in two spatial dimensions and each lattice site of the board contains a single quantum particle. Further, we require that the spatial wavefunctions of these particles are localized to each site so that there is no spatial overlap between the wavefunctions of particles in different sites. From now on we drop the fictitious particle labels, but it should be understood that the underlying wavefunctions for any board state have been (anti) symmetrized.

To determine the quantum state space of the puzzle we start with the solved state and swap particles using the operations in \cref{eq:swaps}. Because these operations are now acting on a quantum state they must be represented by a unitary operator
\begin{equation}\label{eq:SWAPs}
    \mathbb{S}_k \quad \text{for}\,\, k\in \{U,D,R,L\}\,,
\end{equation}
where $U$ denotes a ``Up'' SWAP, and similarly for Down, Right, and Left. Swapping indistinguishable particles leds to a global phase and thus is a trivial operation e.g.
\begin{equation}
\mathbb{S}_U\zero{15pt} = e^{i\phi}\zero{15pt} \quad {\rm and}\quad  \mathbb{S}_D\zero{15pt} = e^{i\phi'}\zero{15pt} \,.
\end{equation}  
Instead if we perform a Left SWAP
\begin{equation}
\mathbb{S}_L\zero{15pt} = \four{15pt} \,,
\end{equation} 
then the puzzle transitions to a new board state. It is easy to show that the new board state is orthogonal to the solved state, i.e. 
\begin{equation}\label{eq:innerproduct}
\puzzIPone{14pt} = \puzzIPtwo{14pt} = 0 \, , 
\end{equation} 
by computing overlaps between the underlying (anti) symmetrized wavefunctions, see \cref{apx:explicit}. By applying sequences of SWAPs to the solved state we find a finite and mutually orthogonal set of basis of states,
\begin{equation}\label{eqn:basis}
\left |\raisebox{-0.42em}{\resizebox{15pt}{!}{\cell{}{}{}{}{gc}{gc}{bc}{bc}}} \!\! \right \rangle, \,\,
\left |\raisebox{-0.42em}{\resizebox{15pt}{!}{\cell{}{}{}{}{bc}{bc}{gc}{gc}}} \!\! \right \rangle, \,\,
\left |\raisebox{-0.42em}{\resizebox{15pt}{!}{\cell{}{}{}{}{bc}{gc}{bc}{gc}}} \!\! \right \rangle, \,\,
\left |\raisebox{-0.42em}{\resizebox{15pt}{!}{\cell{}{}{}{}{gc}{bc}{gc}{bc}}} \!\! \right \rangle, \,\,
\left |\raisebox{-0.42em}{\resizebox{15pt}{!}{\cell{}{}{}{}{bc}{gc}{gc}{bc}}} \!\! \right \rangle, \,\,
\left |\raisebox{-0.42em}{\resizebox{15pt}{!}{\cell{}{}{}{}{gc}{bc}{bc}{gc}}} \!\! \right \rangle  ,
\end{equation}
which correspond to the classical realizations in \cref{eqn:classicalStates}. This means we can alternatively represent our collection of particles in a 6-dimensional or qudit Hilbert space where the SWAPs are permutation or signed permutation matrices depending on whether the particles are bosonic or fermionic. This representation is explored in \cref{apx:slide_qudit}, including matrix representations of the SWAP operations in \cref{eq:SWAPs}. 

Although the puzzle consists of entangled fermions, the SWAP actions we introduced only allow transitions between six distinct and orthogonal basis states. The SWAPs never introduce a superposition between the basis states. So, despite the inherent quantum nature and entanglement, the puzzle is effectively classical.  Similar observations have been made for 
quantum versions of Sudoku, see Refs.~\cite{Nechita_Pillet_2020_sudoq,Paczos2021}.

\section{Escaping the Classical Subspace}\label{sec:quant_perm_puzz}
 
We have argued that although the puzzle operates on quantum principles, it is equivalent to the classical puzzle. Now we propose a minimal addition that will unlock a richer state space that cannot be mapped cleanly onto any classical permutation puzzle. The basic idea is to introduce $\sqrt{\SWAP}$ as a possible action.

All of the operators representing the physical SWAP are unitary and Hermitian so we can exponentiate them to generate fractional SWAPs
\begin{align}
   \mathbb{ F }_k(\theta )=  \exp\big( i\theta \mathbb{S}_k \big) = 
    \cos\big( \theta \big) I + i\sin\big( \theta \big) \mathbb{S}_k,
\end{align}
for $k\in\{U,D,L,R\}$. When $\theta = \pi /2$ we return to the SWAPs up to a global phase of $i$.

We choose to add new actions to our puzzle, $\mathbb{ F }_k(\pi/4)$, that is square root SWAPs of the form 
\begin{align}\label{eq:quantum_moves}
   \HS_k = \sqrt{{\rm SWAP}_k} = \exp \bigg (i \frac{\pi}{4} \mathbb{S}_k  \bigg) = \frac{1}{\sqrt{2}} \big (   I + i\mathbb{S}_k \big ),
\end{align}
where $k\in\{U,D,L,R\}$. This move can be interpreted as the equal superposition of swapping and not swapping two elements and the phase of $i$ is necessary to keep the action unitary. This is different from the related classical action of flipping a coin to decide whether or not each SWAP should be applied. The classical action does not change the number of possible board states. The square root SWAP\ action will make any classical puzzle quantum. For concreteness, in this section and the next, we will consider the case of indistinguishable Fermions.

We can quickly see that this SWAP action will allow our puzzle to reach more board states by applying the square root SWAP to one of our basis states
\begin{equation}
    \HS_R\zero{15pt} = \frac{1}{\sqrt{2}}\left(\zero{15pt}+i\five{15pt}\right).
\end{equation}
A single square root SWAP applied to a basis state gives us access to a state that is an equal superposition of our basis states. Applying two square root SWAPs 
\begin{equation}
    \HS_R\HS_R\zero{15pt} = i\five{15pt},
\end{equation}
is the same as applying a full SWAP, in this case $\mathbb{S}_R$, up to a global phase. Continuing to apply square root SWAPs 
\begin{subequations}
\begin{align}
    \HS_R^3 \zero{15pt} &= -\frac{1}{\sqrt{2}}\left(\zero{15pt}-i\five{15pt}\right)\\     \HS_R^4 \zero{15pt} &= -\zero{15pt} \,,
\end{align}
\end{subequations}
we arrive back to the initial state with a different global phase. Only after applying $\HS_R^8$ will we return to the initial state with the same global phase.

Sequences of noncommuting square root SWAPs on different particles give us states that are no longer equal superpositions. For example with fermions,
\begin{equation}
    \HS_U\HS_R\zero{15pt} = \left(\frac{i}{2}\five{15pt} + \frac{e^{-i\pi/4}}{\sqrt{2}}\zero{15pt}-\frac{1}{2}\two{15pt}\right),
\end{equation}
where the phase $e^{-i\pi/4}$ is a direct result of the fermionic exchange statistics.

A question we now pose is how many distinct states can be reached starting from $\solved$ by applying a sequence of square root SWAPs? Given that the action of the square root SWAP\ is discrete, it may be appealing to make a combinatorial argument to bound the total number of states, however this will not work. It turns out that our puzzle now has access to an infinite number of possible board positions. Showing this uses the group representation of the allowed square root SWAPs. 

Our puzzle has an orthonormal basis of 6 states.
Since the square root SWAP is unitary, the group generated by the set of square root SWAPs, $G(\HS)$, will be a subgroup of $\text{SU}(6)$.
To determine the cardinality of this subgroup, we use the methods outlined in~\citet{Sawicki2017,sawicki_universality_2017}.
They show that a sufficient condition for a subgroup of $\text{SU}(d)$ to be infinite is if the subgroup contains two non-commuting elements, $U_1,U_2$ that are both sufficiently close to a scalar multiple of the identity.
So, if two distinct elements generated by square root SWAPs are sufficiently close in Hilbert-Schmidt distance to the identity and do not commute, then the group generated by the square root SWAPs is infinite.
We numerically implement this method and provide our code in Ref.~\cite{Github}.
Using this code, we show that the set of square root SWAPs on our \nbym{2}{2} puzzle generates an infinite subgroup of $\text{SU}(6)$.
Using another method from~\citet{Sawicki2017} we can also  (numerically) show that the group $G(\HS)$ is a proper subgroup, which means that there are elements of $\text{SU}(6)$ that cannot be generated from any (possibly infinite) sequence of square root SWAPs.

Realistic classical permutation puzzles have a finite state space. In fact, it is unclear what it may mean to solve a permutation with a set of infinite possible states, as typically a solver expects there to be a solving algorithm which is guaranteed to terminate after a bounded number of moves. Fortunately, even though the set of possible board states is infinite, the set of basis states is finite and only contains six elements.
This will allow us to include operations that simplify the state of the puzzle and make the puzzle finitely solvable.

\section{Solving Strategies}\label{sec:solns}

Now that we have a rich structure of allowed operations and possible board positions, will introduce a set of rules to describe scrambling and solving the puzzle.
Our rules are just one possible choice, and various alternative rules could be interesting for future study.
The goal of a permutation puzzle is to use sequences of permutations to go between some randomly chosen scrambled state to some preselected solved state. 
We chose $\solved$ to be the solved state, and we will consider a valid scrambled state to be any state that can reach the solved state via a sequence of square root SWAPs.

The rules of the game are as follows.  Each solver is told the starting scrambled state so that the puzzle can be solved, which is analogous to inspecting the puzzle before solving. Each of the three solvers we describe below have different moves available to them. After applying the moves, the puzzle is presented to a referee who verifies the solution through measurement.
The measurement checks if the puzzle is solved or not. If successful, the puzzle is projected into the solved state; if not, an outcome-dependent unitary is applied to reset the puzzle to the original scramble and the solver must start again. To describe the measurement ($\mathbb{M}$) we introduce the projectors
\begin{equation}
\begin{aligned}
     \Pi_s &= \zero{15pt}\zerod{15pt}\\
     \Pi_{1} &=  \one{15pt}\oned{15pt}\\
     \vdots\,\,\, &\\
     \Pi_{5} &=\five{15pt}\fived{15pt}.
\end{aligned}
\end{equation}
After projection, the board is returned to the scrambled state with the outcome-dependent unitary $U_i$ for each outcome $i\neq s$, that is $U_i\Pi_{i}$. Finally, every solver move and verification measurement is counted and penalized as a move -- we will give more details on the cost function shortly. Solvers aim to minimize the expected move count by leveraging classical computation to optimize their strategies and maximize success probability.

A ``quantum solver'' can use the square root SWAPs, $\{\mathbb{H}_k\}$, to move between the scramble state and the solved state. That is their actions are $A_{\rm quan}= \{\HS_U, \HS_D, \HS_L,\HS_R, \mathbb{M}\}$.\footnote{Technically the measurement is the referee's action, but the decision to hand the puzzle to the referee will count as an action for the solver.} 
A quantum solver can input the square root SWAPs into a unitary synthesis protocol to construct a (non-unique) sequence of moves to solve the puzzle. Thus, a quantum solver can return any scrambled state to the solved state, but may require many square root SWAPs and access to unlimited classical computation for certain scrambles.

 A ``classical solver'' is restricted to using SWAPs to try and solve the puzzle. The classical SWAPs can permute the amplitudes on each of the six basis states but can never concentrate amplitude. Consequently, the classical solver can only deterministically solve the puzzle if the scrambled state is one of the six classical basis states. Thus the measurement becomes a valuable resource for the classical solver. In summary the classical solvers actions are $A_{\rm class}= \{\mathbb{S}_U, \mathbb{S}_D, \mathbb{S}_L, \mathbb{S}_R, \mathbb{M}\}$. 
The obvious stratergy for the classical solver is to move the amplitude with the larget magnitude onto the solved state.  For example, consider the board state below and a Right SWAP which moves the largest amplitude onto the solved state
\begin{equation}
\mathbb{S}_R \left(\sqrt{\frac{3}{4}}\five{15pt} +\sqrt{\frac{1}{4}}\two{15pt}\right) \rightarrow \sqrt{\frac{3}{4}}\zero{15pt} -\sqrt{\frac{1}{4}}\two{15pt}.
\end{equation}
After the Right SWAP the probability for projecting into the solved state is $\Pr(s) = 0.75$.

The final solver can use either square root SWAPs or full SWAPS, so we call it a ``combined solver''. In summary the combined quantum--classical solvers actions are $A_C= A_{\rm class}\cup A_{\rm quant}$. The key difference between the combined solver and quantum solver is the cost of classical SWAPs. While the quantum solver can always generate a SWAP in two moves (eg. $\mathbb{H}_k\mathbb{H}_k \propto \mathbb{S}_k$), the combined solver only takes one move to accomplish the same action.

Our chosen cost function evaluates solver performance by expected move count required to solve the puzzle. It accounts for the number of moves taken to reach a high-probability state and the measurement step, which may require resets if unsuccessful.
Let the random variable $X_\varphi$ be the number of moves required to solve the puzzle using the candidate state $|\varphi\rangle$. 
Using a candidate state means the solver has taken $M_\varphi$ actions to arrive at $|\varphi\rangle$ from the scramble and then measures.
If the measurement fails and the puzzle is reset, then the solver repeats this procedure. 
Because of this resetting behavior the random variable follows a geometric distribution: $X = (M_\varphi+1)\text{Geo}(P_\varphi)$, where the $M_\varphi+1$ comes from the additional one move required to measure and $P_\varphi$ is the probability the board state is projected into the solved state. 
The mean of this random variable is $E[X] = (M_\varphi+1)/P_\varphi$. This will serve as the cost function that the solvers will minimize with their classical computation.

\begin{figure}[!th]
    \centering
    \includegraphics[width=\columnwidth]{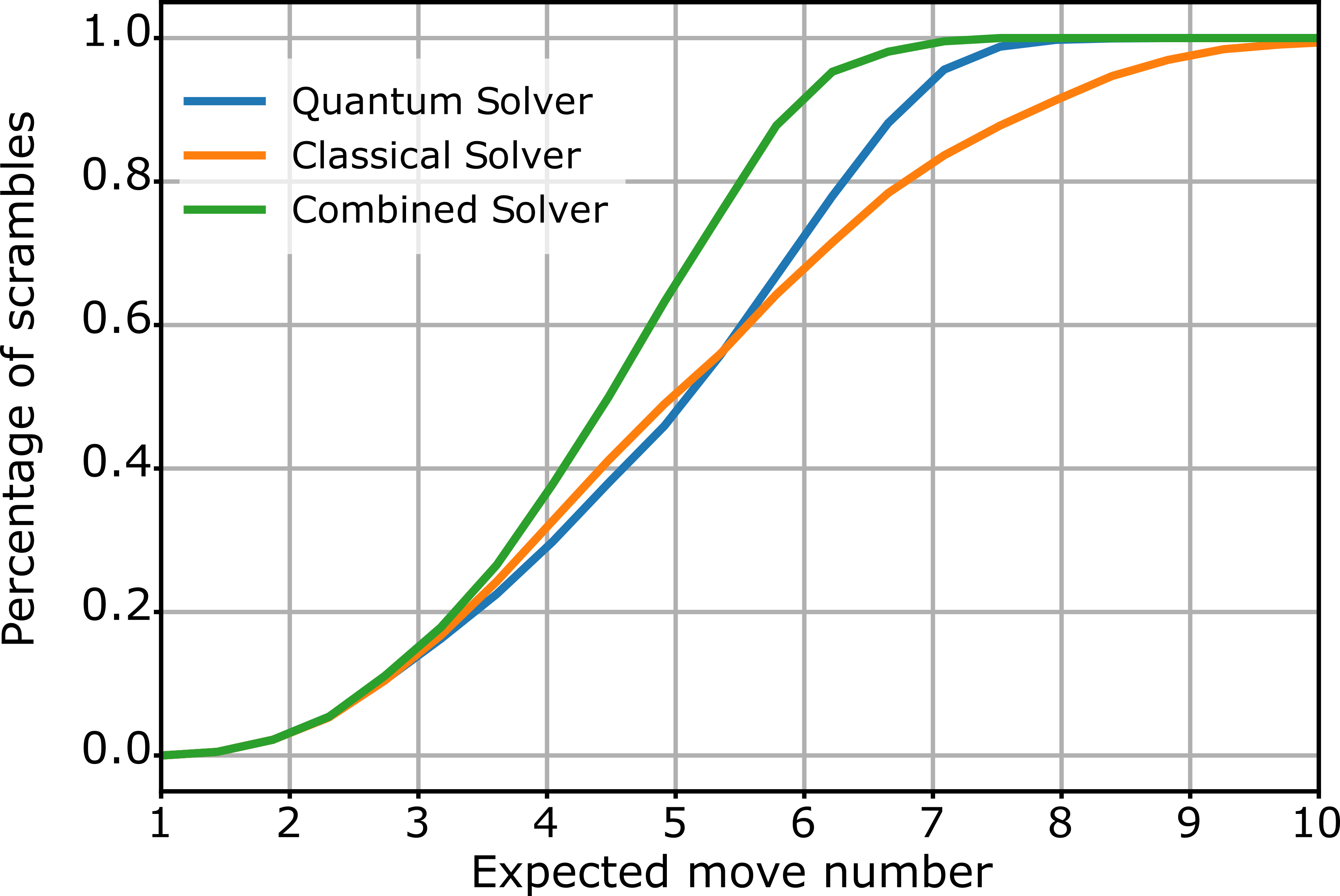}
    \caption{ The numerically evaluated cumulative distribution function of each solvers expected move count. This was numerically evaluated using 2000 random scrambles (average scramble length was 350 square root SWAPs) with each scramble being solved optimally. The classical solver can move amplitude quicker than the quantum solver allowing it to solve a greater portion of scrambles in less than 5 moves. But the classical solver's inability to concentrate amplitude leads to some scrambles requiring many moves to solve. The quantum solver almost never encounters a scramble that takes more than 8 moves to solve on average.  The combined solver is strictly superior to both the quantum and classical solver because it contains both those strategies. Overall the classical solver is worst, taking 5.88 moves on average to solve a scramble on average. The quantum solver is about half a move better, taking 5.32 moves on average. Finally the combined solver is the best, taking just 4.77 moves on average.
    }\label{fig:solving}
\end{figure}

In \cref{fig:solving} we simulate the performance of three solvers: the quantum solver with access to square root SWAPs, the classical solver with access to SWAPs, and a combined solver that can do both square root SWAPs and classical SWAPs each with just a single move. The classical solver can solve some scrambles much faster than the quantum solver, but the lack of versatility in the classical strategy leads to some scrambles taking many moves to solve. The combined solver mitigates this early advantage that comes from the faster SWAP action, but does not suffer from difficult scrambles. Overall, the classical solver does the worst on average with an average move count of 5.88 moves. The quantum solver is noticeably better taking only $5.32$ moves on average. And, as expected, the combined solver is much better at solving the puzzle in $4.77$ moves on average.
While puzzles are typically single-player challenges and not directly related to multi-player games, there are similarities in how quantum agents can outperform classical strategies. For example, see Refs.~\cite{Eisert_etal_1999, GutoskiWatrous_2007} from the quantum games literature.

\section{Universality}\label{sec:universality}

Our $2\times 2$ puzzle corresponds to a six level qudit as explained in \cref{sec:perm_puzz}.
Although the state space of our puzzle is infinite, not every linear combination of basis states is achievable with just square root SWAPs starting from \solved. We describe some of these unreachable states in \cref{app:invariant}. We now explore adding some additional actions to the solver's repertoire so that all linear combinations become achievable.
In quantum computing, a universal gate set enables one to generate an arbitrary approximation to any linear combination of basis states starting from any initial state.
Specifically, if the set of operations generated by a gate set is dense in $\text{SU}(d)$, then it is said to be a universal gate set for a $d$-level qudit.

It might be helpful to first consider why the set of square root SWAPs is not universal. We determined this by applying the procedure introduced in~\cite{Sawicki2017}, which checks two requirements.  The first requirement is that the group generated from the set of square root SWAPs is infinite. The second requirement is that the set of matrices that commute with the adjoint representation of the generators is just scalar multiples of identity.
Our puzzle passes the first check but not the second. A direct result of our puzzle having two blue particles and two green particles is that the operation that exchanges the color of the particles commutes with all square root SWAPs, see \cref{app:centralizer} for more details.

To achieve universality, we can add additional gates to supplement the square root SWAPs. One simple additional gate is the
$\mathbb{P}_6$ gate
\begin{equation}\label{eq:P_6}
\mathbb{P}_6 = \text{diag}\left(1,1,1,1,1,-1\right)
\end{equation}
which applies a $\pi$ phase to one of the classical basis states, but does not permute any of the states amongst themselves. This gate does not have a clear interpretation in terms of solving the puzzle, as it is not related to an actual permutation of our puzzle pieces. 
If we take the original gate set and include the $\mathbb{P}_6$ gate, i.e. $ \{\HS_U, \HS_B, \HS_L, \HS_R\} \bigcup \{\mathbb{P}_6 \}$, and run the numerical proof, we find that this new gate set passes the universality test.
The code implementing these tests is available in Ref.~\cite{Github}.

It is also possible to achieve universality by modifying the building blocks of our puzzle. We now allow different colored particles of our puzzle to also have different exchange statistics, e.g. green bosons and blue fermions. In this case, a color SWAP also swaps exchange statistics, leading to observable changes in the puzzle. Because bosons and fermions pick up a different phase when exchanged, the relative phase between basis states is no longer connected to the color symmetry. This relative phase difference serves a similar function to the phase introduced by the $\mathbb{P}_6$ gate. 
This phase difference, in addition to adding square root diagonal SWAPs or full diagonal SWAPs, is enough to make the gate set universal on this new puzzle. To verify this, we change the gate set that we input into the universality-checking algorithm so that this different phase is reflected.
In fact, this is equivalent to replacing one of the sets of fermions with empty sites. We show in \cref{app:vacuumswaps} that these sites would have positive exchange statistics and are trivially indistinguishable, so they allow the puzzle to be universal.

\section{Hilbert space dimension of larger board sizes}\label{sec:board_size}

We have so far focused on the relatively small \nbym{2}{2} board that has $4!/(2! \times 2!) = 6$ unique classical board states which we mapped to a $6$ dimensional qudit Hilbert space. We now compute the qudit dimension of boards with more positions and colors. We will then briefly argue that the quantum advantage of the puzzle grows with board size.

One simple generalization of our puzzle is to consider two colors of particles in a larger \nbym{n}{n} grid of board positions. This will give us a total of $n^2!$ arrangements of the board positions, of which many will be indistinguishable. For the simplest case where $n$ is even and thus we have $n^2/2$ particles of each color we can map this puzzle to an effective qudit dimension of
\begin{equation}\label{eq:n_by_n_2color_even}
d_{\rm even} = \frac{n^2!}{(n^2/2)! \times (n^2/2)!} = \binom{n^2}{n^2/2} \sim \sqrt{\frac{2}{\pi}} \frac{2^{n^2}}{n} \,,
\end{equation}
where we have divided through by the $(n^2/2)!$ arangements of each color of indistinguishable particles. Interestingly, for even $n$, the sequence of numbers given by $d_{\rm even}$ is a well known integer sequence~\cite{A081623}.
When $n$ is odd, we can simply assume that there is a single green fermion added, and then the asymptotic effective qudit dimension is identical to \cref{eq:n_by_n_2color_even}.
 
 One can also consider \nbym{n}{n} boards, with $n$ colors of particles. In this case it does not matter if $n$ is odd or even, because each color of particle has $n!$ indistinguishable configurations and we have $n$ colors. The equivalent qudit Hilbert space dimension grows as~\cite{A034841}
\begin{equation}\label{eq:n_by_n_ncolor}
d = \frac{n^2!}{\underbrace{n! \times n! \ldots \times n!}_{n}} =
\frac{n^2!}{(n!)^n}  \sim n^{n^2-\frac{n}{2}+1} (2 \pi )^{\frac{1-n}{2}}\, ,
\end{equation}
which grows much faster than \cref{eq:n_by_n_2color_even}.

For non-square puzzles it is useful to define a puzzle to be \emph{connected} if for any two board positions there exist a series of nearest neighbor SWAPs that connect the two positions. Then we can consider a connected board with $N$ total board positions, that is populated with $m$ colors of particles with particle numbers $n_1,n_2,\dots n_m$ such that $\sum n_i = N$.  For such a board, the total number of board states or the effective qudit Hilbert space dimension is
\begin{equation}\label{eqn:generalboard}
   d =  \frac{N!}{n_1!n_2!n_3!\dots n_m!} = \binom{N}{n_1,n_2,\dots,n_m}\, ,
\end{equation}
where the right-hand side is a multinomial coefficient. Clearly \cref{eq:n_by_n_2color_even} and \cref{eq:n_by_n_ncolor} are special cases of \cref{eqn:generalboard}. 

Let us briefly consider how the quantum advantage seen in \cref{sec:solns} might scale with board size. Providing an exact description is challenging, as it is known that finding the optimal solution to the classical \nbym{n}{n} puzzle is NP-hard~\cite{RatnerWarmuth1986}. Instead, we provide a conjecture based on our observations in the \nbym{2}{2} puzzle.

We conjecture that the advantage the combined solver and quantum solver have over the classical solver will increase with Hilbert space dimension. This conjecture is based on the observation that the classical solver struggles to solve scrambles with less concentrated amplitudes. When changing puzzle geometries, i.e the total number of particles and connectivity, it is less clear whether quantum advantage should increase or decrease. 
In \cref{app:quant_adv} we provide some preliminary evidence for this conjecture by investigating puzzles with increasing Hilbert space dimension for two classes of puzzle, one where quantum advantage is present, and one where it is not.

We can also directly calculate the universality of many different puzzle shapes. Because the universality proof is numerical, we cannot prove the universality of the general case, but we outline our findings on small puzzles in \cref{app:universal}. We conjecture that no puzzle with only bosons or fermions is universal with just square root SWAPs. We further conjecture that the addition of a $\mathbb{P}_N$ gate of the form given by \cref{eq:P_6} makes almost every puzzle universal. For disconnected puzzles, notions of SWAPs between the disconnected subspaces of the puzzle can be made universal~\cite{vanmeter2021}.

\section{A quantum cube puzzle}\label{sec:cube}
Previously, we have considered puzzles in two dimensions with nearest-neighbor SWAPs. This does not capture the complexity of puzzles like the Rubik's cube. Here, we introduce a quantum version of the $2\times2\times1$ Rubik cube shown in \cref{fig:2x2x1_cube} (a). The visible colored faces of the puzzle pieces are replaced with identical particles. We flatten the 3D puzzle into 2D, as depicted in \cref{fig:2x2x1_cube}(b), and consider permutations of the particles consistent with rotational configurations. In \cref{app:3D_cube_rotations}, we briefly outline the full three-dimensional modeling of the puzzle. Surprisingly, in both cases the puzzle can be mapped onto the \nbym{2}{2} slide puzzle considered above.

\begin{figure}[th!]
\flushleft
{\hspace{0.6cm}\footnotesize (a) \hspace{3.5cm} (b)}\\ 
\vspace{-0.25cm}
\centering
\includestandalone[width=0.33\columnwidth]{cube_2x2x1} \hspace{40pt}
\includestandalone[width=0.33\columnwidth]{c0_no_lab}
    \caption{A three dimensional permutation puzzle. The solved state of the $2\times 2\times 1$ Rubik's cube puzzle is depicted in (a). In (b) we represent the solved state as a 2D plane, including the pieces that are not visible in (a).
    }\label{fig:2x2x1_cube}
\end{figure}

The $2\times 2\times 1$ cube has six unique states, see \cref{fig:cube_states}, up to an orientation assumption. In this section, we assume that the solution of the cube depends only on the relative arrangement of colors, not on the orientation of the cube.\footnote{This differs from our treatment of the slide puzzle in \cref{sec:perm_puzz}. In 3D puzzles like the Rubik's cube, the puzzle's orientation is irrelevant, and ignoring this would unnecessarily expand the state space.}
Any scramble of this puzzle can be solved in at most three moves, i.e. ``God's number'' is three.

The primitive moves on the puzzle are $\pi$ rotations of one of the $2\times 1\times 1$ portions of the puzzle. 
The possible moves on the cube are rotations of the Left, Right, Up, Down which are denoted $\{U,D,R,L\}$~\cite{Signmaster81}. In this puzzle the moves $\{U,R\}$ are sufficient to generate and solve all possible cube states, due to our orientation assumption. Specifically starting with the solved state, the moves
\begin{align}\label{eq:class_moves}
U,\, R,\, UR,\, R\,U,\, UR\,U,
\end{align}
generate the board states in \cref{fig:cube_states}. The choice of $\{U,R\}$ is not unique, any two non-commuting or intersecting rotations generate all board states.

It is possible to go through the same, albeit more complicated, procedure in the slide puzzle to correctly represent the underlying identical particles as we did in \cref{apx:explicit},  see \cref{app:3D_cube_rotations}.
As the flattened board state is too unwieldy to fit inside a ket, we instead prescribe a canonical ordering of the states with the following kets
\begin{equation}
    \ket{0}, \ket{1}, \ket{2}, \ket{3}, \ket{4}, \ket{5} \,.
\end{equation}
In \cref{fig:cube_states}, we give the mapping between these kets and the cube state.

\begin{figure}[ht!]
\includestandalone[width=0.28\columnwidth]{c0} 
\hspace{8pt}
\includestandalone[width=0.28\columnwidth]{c1}
\hspace{8pt}
\includestandalone[width=0.28\columnwidth]{c2}\\
\vspace{10pt}
\includestandalone[width=0.28\columnwidth]{c3}
\hspace{8pt}
\includestandalone[width=0.28\columnwidth]{c4}
\hspace{8pt}
\includestandalone[width=0.28\columnwidth]{c5}
\caption{All the classical board states. The first state is the solved state, the others can be obtained by $U, R, U\!R, RU, U\!RU$. The kets illustrate the correspondence between the classical board states and our quantum basis. In the first state, $\ket{0}$, the thick black lines outline the colored faces of a single physical puzzle piece—a $1\times1\times1$ block known as a ``cubie.'' There are four cubies in this puzzle. The remaining board states are rotated and permuted versions of the original cubies.
}\label{fig:cube_states}
\end{figure}

In the flattened form, it is difficult to see how the rotations change the state, as the natural moves are no longer a SWAP of two identical particles. The rotations effectively swap several particles.  For this reason, it is useful to think of the moves as permutations. Thus transitioning between board states is a permutation of several particles so we denote the permutations by
\begin{align}
     \mathbb{P}_U,\quad {\rm and}\quad \mathbb{P}_R \, ,
\end{align}
which are self inverse so $\mathbb{P}_{U'}=\mathbb{P}_{U}$ and $\mathbb{P}_{R'} = \mathbb{P}_{R}$. Acting these permutations on the board indeed transitions the board to a new state, e.g. $\mathbb{P}_U\ket{0} = \ket{1}$ and $\mathbb{P}_R\ket{0} = \ket{2}$.
In \cref{apx:cube_qudit}, we give the matrix representation of the permutation matrices. As with the slide puzzle, although the underlying bosons are entangled, the permutations merely map between six orthogonal states. Consequently, the puzzle is equivalent to the classical $2 \times 2 \times 1$ puzzle.

The flattened form of the puzzle is a natural representation, as four pieces that physically make up the puzzle can be isolated -- see \cref{fig:cube_states}. Theses four pieces are the $1\times 1\times1$  blocks that each have four particles and are called  ``cubies''. For this representation, solving the puzzle depends only on the relative arrangement of the cubies, not the puzzle's orientation. Because the cubies are distinguishable, there are, naively, $4! = 24$ different cubie combinations. Each rotation swaps and reorients two cubies. In this specific puzzle, cubie orientation is completely dependent on position, so we can neglect that degree of freedom. Then the $4! = 24$ different cubie combinations can be reduced by removing states that are equivalent up to the orientation of the puzzle giving $4!/4 = 6$ unique states. $\mathbb{P_U}$ and $\mathbb{P}_R$ are now directly interpretable as SWAPs of the four cubies, which is analogous to the operations on the 2D puzzle. The key difference is the distinguishability of the cubies and the restricted set of allowed SWAPs.\footnote{  The puzzle can also be represented as a $3 \times 1$ puzzle with three distinguishable particles.
  In \cref{fig:cube_states}, the bottom-left cubie remains fixed. Ignoring it and stacking the rest gives a $3 \times 1$ puzzle with $3! = 6$ states. The permutation $\mathbb{P}_U$ swaps the top and middle cubies, while $\mathbb{P}_R$ swaps the middle and bottom cubies.}

As before, we choose to add a new action that a solver can take, the square root of a permutation
\begin{equation}\label{eq:quantum_perm}
   \mathbb{Q}_k = \sqrt{{\rm PERM}_k} = \exp[i (\pi/2) \mathbb{P}_k /2] = \frac{1}{\sqrt{2}} \big (   I + i\mathbb{P}_k \big ),
\end{equation}
where $k\in\{ U, R\}$. Acting $\mathbb{Q}_k$ on the solved state produces a superposition as you would expect e.g.
\begin{subequations}
\begin{align}
\mathbb{Q}_U \ket{0} &= \frac{1}{\sqrt{2}}(\ket{0} + i\ket{1})\\
\mathbb{Q}_R \ket{0} &= \frac{1}{\sqrt{2}}(\ket{0} + i\ket{2})\,. 
\end{align}
\end{subequations}
As before, sequences of noncommutting square root permutations lead to states that are no longer equal superpositions. And thus we have created a quantum version of the $2\times 2\times 1$ Rubik's cube with a state space that is analogous to our \nbym{2}{2} slide puzzle.

 In \cref{app:3D_cube_rotations}, we give a treatment of the puzzle in 3D using rotations. Ultimately we arrive back to a puzzle with six unique states and add a square root SWAP\ to make the puzzle more quantum.

\section{Conclusion}\label{sec:concl}

We have introduced a quantum generalization of permutation puzzles which does not have a simple classical analog.
The core idea was to represent the different colored pieces of a permutation puzzle with sets of identical quantum particles and use a quantum move that created superpositions between classically allowed states.
This puzzle can be solved deterministically with quantum moves or, with the addition of measurements, be solved nondeterministically with just classical moves.

With superpositions, the number of unique allowed states of the puzzle is infinite, unlike common permutation puzzles from toy stores. While this bigger state space allows for a richer puzzle structure we can also consider making finite quantum puzzles. One way to do this would be, for the case of e.g. identical fermions, to consider the allowed moves to be a subset of fermionic Gaussian unitaries~\cite{knill2001fermionic,BravyiKitaev2002,JozsaMiyake2008} that correspond to discrete Clifford elements~\cite{ZhaoRubinMiyake2021}. Such a restriction would result in a discrete (but large) state space.
Since the state space and set of actions are both finite, these versions of the puzzles could be mapped onto classical permutation puzzles, albeit ones with very strange geometries. {These finite puzzle's are more similar to other notions of quantum puzzles such as quantum sudoku, which are characterized by the cardinality of allowed states~\cite{Paczos2021}.}

There are several questions that we think naturally arise from our discussion here. First, there is the question of implementation. Arrays of ultracold atoms in optical lattices~\cite{Endres2016,Ebadi2021,Kaufman2021} might be a natural platform because confinement and swapping of both fermionic and bosonic atoms has been demonstrated. Moreover recent experiments have demonstrated confinement of multiple species of atoms~\cite{Singh2022,Anand2024,Beterov2015}. Quantum dot plaquettes~\cite{ButerakosDasSarma2019} could make an appealing experimental platform given recent experimental demonstrations~\cite{Dehollain2020}.

The second question we highlight is: what is the ``puzzle group'' generated by the square root SWAPs? The group is not $\text{SU}(6)$ as our actions are not universal for $\text{SU}(6)$. The actions may be universal on $\text{SU}(k)$ for some $k<6$.

Finally, there are many more permutation puzzles (classical and quantum) than we have described here. 
On the quantum side one could imagine replacing the fermions and bosons with Anyons~\cite{AnyonsWilczek82}. On the classical side we think our examples above provide a recipe to quantize permutation puzzles: 
\begin{enumerate}[font=\itshape]
    \item \textit{Replace the structure of the puzzle with sets of identical particles.} If the quantum particles are appropriately indistinguishable then the state space of the particles will match the state space of the classical puzzle. 
    \item  \textit{The classical primitive moves (swaps, permutations, rotations) can easily be quantized as permutation matrices acting on the state space.} These permutation matrices are unitary and naturally act on the Hilbert space of the quantum puzzle. 
    \item \textit{To achieve a non-classical puzzle one adds a quantum move.} We have argued that unitaries that are generated by the exponential of permutations are a natural candidate.
\end{enumerate}

\noindent {\em Acknowledgments:} The authors acknowledge helpful discussions with Todd Brun, Shawn Geller, Joshua Grochow, Kyle Gulshen, and Robert Smith. NL, AK, and JC were supported by the University of Colorado Boulder. MTG would like to thank Sharon Anderson and the CU Boulder Summer Program for Undergraduate Research (CU SPUR) for their support.

\bibliography{refs.bib}
\vspace{2em}

\newpage\newpage

\appendix

\section{Inner product in explicit Fermion representation}\label{apx:explicit}
The permutation puzzle has four physical locations:
\begin{equation}
\resizebox{0.3\columnwidth}{!}{
\begin{tikzpicture}
\node[rectangle, minimum width =16pt, minimum height =16pt ] at (-17pt,17pt)    {\small Loc. 1};
\node[rectangle, minimum width =16pt, minimum height =16pt ] at (17pt,17pt)    {\small Loc. 2};
\node[rectangle, minimum width =16pt, minimum height =16pt ] at (-17pt,-17pt)    {\small Loc. 3};
\node[rectangle, minimum width =16pt, minimum height =16pt ] at (17pt,-17pt)    {\small Loc. 4};
\draw[step=34pt,black,thin] (-34pt,-34pt) grid (34pt,34pt);
 \end{tikzpicture}} \, .
\end{equation}
To formalize our diagrammatic representation, we want to represent these locations (or positions) with a tensor product structure as
\begin{equation}\label{eq:hilbert_struct}
    \ket{\text{location 1, location 2, location 3, location 4}}\, .
\end{equation}
Any of these locations can hold a green or a blue fermion e.g. $\ket{\text{location 1}}$ would decompose further as 
\begin{equation}\label{eq:color_tensor}
 \ket{\text{location 1}}= \ket{\text{green}}\otimes \ket{\text{blue}}   \, .
\end{equation} 
Thus one can even write down fermionic creation operators to transition from the vacuum to a particular state. In the main text we represented the solved state as
\begin{alignat}{5}\label{eq:sovled_app}
 \left |\raisebox{-0.42em}{\resizebox{15pt}{!}{\cell{\phantom{1}}{\phantom{1}}{\phantom{1}}{\phantom{1}}{gc}{gc}{bc}{bc}}} \!\! \right \rangle &{}\propto{}&
       \left |\raisebox{-0.42em}{\resizebox{15pt}{!}{\cell{1}{2}{1}{2}{gc}{gc}{bc}{bc}}} \!\! \right \rangle  
&{}-{}&\left |\raisebox{-0.42em}{\resizebox{15pt}{!}{\cell{2}{1}{1}{2}{gc}{gc}{bc}{bc}}} \! \right \rangle 
&{}-{}&\left |\raisebox{-0.42em}{\resizebox{15pt}{!}{\cell{1}{2}{2}{1}{gc}{gc}{bc}{bc}}} \!\! \right \rangle  
&{}+{}&\left |\raisebox{-0.42em}{\resizebox{15pt}{!}{\cell{2}{1}{2}{1}{gc}{gc}{bc}{bc}}} \! \right \rangle 
\end{alignat}
Let's rewrite this state with respect to the tensor product structure in \cref{eq:hilbert_struct} and leave the color tensor product in \cref{eq:color_tensor} implicit:
\begin{widetext}
\begin{align}\label{eq:zero_app}
\left |\raisebox{-0.42em}{\resizebox{15pt}{!}{\cell{\phantom{1}}{\phantom{1}}{\phantom{1}}{\phantom{1}}{gc}{gc}{bc}{bc}}} \!\! \right \rangle &\propto\ket{\g{1},\g{2},\blu{1},\blu{2}} -\ket{\g{2},\g{1},\blu{1},\blu{2}}-\ket{\g{1},\g{2},\blu{2},\blu{1}}+\ket{\g{2},\g{1},\blu{2},\blu{1}} \,,
\end{align}
this gives meaning to the diagrammatic representation in \cref{eq:sovled_app}. As another example lets consider another the board state
\begin{align}\label{eq:four_app}
\left |\raisebox{-0.42em}{\resizebox{15pt}{!}{\cell{\phantom{1}}{\phantom{1}}{\phantom{1}}{\phantom{1}}{bc}{gc}{gc}{bc}}} \!\! \right \rangle 
 &\propto  \ket{\blu{1},\g{1},\g{2},\blu{2}}  - \ket{\blu{2},\g{1},\g{2},\blu{1}} - \ket{\blu{1},\g{2},\g{1},\blu{2}} + \ket{\blu{2},\g{2},\g{1},\blu{1}}    \,. 
\end{align}
We can compute the inner product between these term by term. However it is easy to compute one term explicitly
\begin{align}
\ip{\blu{1},\g{1},\g{2},\blu{2}}{\g{1},\g{2},\blu{1},\blu{2}}
 &\propto  \ip{0,\blu{1}}{\g{1},0}\otimes  \ip{\g{1},0}{\g{2},0} \otimes \ip{\g{2},0}{0,\blu{1}} \otimes \ip{0,\blu{2}}{0,\blu{2}}  = 0\, ,
\end{align}
where we separated the tensor product by locations, and the commas separate color.
\end{widetext}
We can do this for all such terms in \cref{eq:zero_app} and \cref{eq:four_app} to show that 
\begin{equation*}
\puzzIPone{14pt} = \puzzIPtwo{14pt} = 0 \, , 
\end{equation*} 
which is \cref{eq:innerproduct}.

\section{slide puzzle Qudit mapping}\label{apx:slide_qudit}
In the main text, see \cref{eq:board0symm}, we defined board states where the underlying identical particles were (anti) symmetrized
\begin{alignat*}{5}
 \left |\raisebox{-0.42em}{\resizebox{15pt}{!}{\cell{\phantom{1}}{\phantom{1}}{\phantom{1}}{\phantom{1}}{gc}{gc}{bc}{bc}}} \!\! \right \rangle &{}\propto{}&
       \left |\raisebox{-0.42em}{\resizebox{15pt}{!}{\cell{1}{2}{1}{2}{gc}{gc}{bc}{bc}}} \!\! \right \rangle  
&{}-{}&\left |\raisebox{-0.42em}{\resizebox{15pt}{!}{\cell{2}{1}{1}{2}{gc}{gc}{bc}{bc}}} \! \right \rangle 
&{}-{}&\left |\raisebox{-0.42em}{\resizebox{15pt}{!}{\cell{1}{2}{2}{1}{gc}{gc}{bc}{bc}}} \!\! \right \rangle  
&{}+{}&\left |\raisebox{-0.42em}{\resizebox{15pt}{!}{\cell{2}{1}{2}{1}{gc}{gc}{bc}{bc}}} \! \right \rangle \,\, (\text{fermionic})\\
 \left |\raisebox{-0.42em}{\resizebox{15pt}{!}{\cell{\phantom{1}}{\phantom{1}}{\phantom{1}}{\phantom{1}}{gc}{gc}{bc}{bc}}} \!\! \right \rangle &{}\propto{}&
       \left |\raisebox{-0.42em}{\resizebox{15pt}{!}{\cell{1}{2}{1}{2}{gc}{gc}{bc}{bc}}} \!\! \right \rangle  
&{}+{}&\left |\raisebox{-0.42em}{\resizebox{15pt}{!}{\cell{2}{1}{1}{2}{gc}{gc}{bc}{bc}}} \! \right \rangle 
&{}+{}&\left |\raisebox{-0.42em}{\resizebox{15pt}{!}{\cell{1}{2}{2}{1}{gc}{gc}{bc}{bc}}} \!\! \right \rangle  
&{}+{}&\left |\raisebox{-0.42em}{\resizebox{15pt}{!}{\cell{2}{1}{2}{1}{gc}{gc}{bc}{bc}}} \! \right \rangle . \,\, (\text{bosonic}) \,\,\,\, 
\end{alignat*}
By performing the SWAP operations we can list all possible board states
\begin{equation*}
\left |\raisebox{-0.42em}{\resizebox{15pt}{!}{\cell{}{}{}{}{gc}{gc}{bc}{bc}}} \!\! \right \rangle, \,\,
\left |\raisebox{-0.42em}{\resizebox{15pt}{!}{\cell{}{}{}{}{bc}{bc}{gc}{gc}}} \!\! \right \rangle, \,\,
\left |\raisebox{-0.42em}{\resizebox{15pt}{!}{\cell{}{}{}{}{bc}{gc}{bc}{gc}}} \!\! \right \rangle, \,\,
\left |\raisebox{-0.42em}{\resizebox{15pt}{!}{\cell{}{}{}{}{gc}{bc}{gc}{bc}}} \!\! \right \rangle, \,\,
\left |\raisebox{-0.42em}{\resizebox{15pt}{!}{\cell{}{}{}{}{bc}{gc}{gc}{bc}}} \!\! \right \rangle, \,\,
\left |\raisebox{-0.42em}{\resizebox{15pt}{!}{\cell{}{}{}{}{gc}{bc}{bc}{gc}}} \!\! \right \rangle  .
\end{equation*}
Then by computing inner products of the underlying identical particle one can show that the six states are orthogonal. Thus these states correspond to a 6 dimensional qudit which we represent by the column vectors
\begin{subequations}
\begin{align}
\ket{0} &= \begin{pmatrix}
 1 & 0 & 0 & 0 & 0 & 0 
\end{pmatrix}^T\\
\ket{1} &= \begin{pmatrix}
 0 & 1 & 0 & 0 & 0 & 0 
\end{pmatrix}^T\\
\ket{2} &= \begin{pmatrix}
 0 & 0 & 1 & 0 & 0 & 0 
\end{pmatrix}^T\\
\ket{3} &= \begin{pmatrix}
 0 & 0 & 0 & 1 & 0 & 0 
\end{pmatrix}^T\\
\ket{4} &= \begin{pmatrix}
 0 & 0 & 0 & 0 & 1 & 0 
\end{pmatrix}^T\\
\ket{5} &=\begin{pmatrix}
 0 & 0 & 0 & 0 & 0 & 1 
\end{pmatrix}^T .
\end{align}
\end{subequations}

For the moment let's focus on the fermionic case. To determine the 6 dimensional unitaries that correspond  to the SWAP operations we simply act with a particular SWAP on the basis of board states.
This will represent the unitary in the board basis. For example, consider the Up SWAP operation acting on the solved board state
\begin{equation}
\mathbb{S}_U\solved = -1 \solved\, .
\end{equation}
This indicates that the unitary has a $-1$ in the top left position. Acting the Up SWAP on the remaining board states gives
\begin{subequations}
\begin{align}
\mathbb{S}_U\one{15pt} &= -1\one{15pt} \\
\mathbb{S}_U\two{15pt} &= +1\five{15pt} \\
\mathbb{S}_U\three{15pt} &= +1\four{15pt} \\
\mathbb{S}_U\four{15pt} &= +1\three{15pt} \\
\mathbb{S}_U\five{15pt} &= +1\two{15pt} \,.
\end{align}
\end{subequations}
If we collect all of this information together in a matrix we find the Unitary matrix that corresponds to applying a Up SWAP is
\begin{equation}
\mathbb{S}_U = \begin{pmatrix*}[r]
-1 &   0 &  \phm 0 &  \phm 0 &  \phm 0 &  \phm 0 \\
 0 &  -1 &  0 &  0 &  0 &  0 \\
 0 &   0 &  0 &  0 &  0 &  1 \\
 0 &   0 &  0 &  0 &  1 &  0 \\
 0 &   0 &  0 &  1 &  0 &  0 \\
 0 &   0 &  1 &  0 &  0 &  0 
\end{pmatrix*}\, ,
\end{equation}
which is a signed permutation matrix.

If we repeat this process for the remaining SWAP operations we find
\begin{equation}
\mathbb{S}_D = \begin{pmatrix*}[r]
-1 &   0 &  \phm 0 &  \phm 0 &  \phm 0 &  \phm 0 \\
 0 &  -1 &  0 &  0 &  0 &  0 \\
 0 &   0 &  0 &  0 &  1 &  0 \\
 0 &   0 &  0 &  0 &  0 &  1 \\
 0 &   0 &  1 &  0 &  0 &  0 \\
 0 &   0 &  0 &  1 &  0 &  0 
\end{pmatrix*}\, ,
\end{equation}
and 
\begin{equation}
\mathbb{S}_R = \begin{pmatrix*}[r]
 \phm 0 &   \phm 0 &  0 &  0 &  \phm 1 &  \phm 0 \\
 0 &   0 &  0 &  0 &  0 &  1 \\
 0 &   0 & -1 &  0 &  0 &  0 \\
 0 &   0 &  0 & -1 &  0 &  0 \\
 1 &   0 &  0 &  0 &  0 &  0 \\
 0 &   1 &  0 &  0 &  0 &  0 
\end{pmatrix*}
\end{equation}
and 
\begin{equation}
\mathbb{S}_L = \begin{pmatrix*}[r]
\phm 0 &  \phm 0 &  0 &  0 &  \phm 0 &  \phm 1 \\
 0 &   0 &  0 &  0 &  1 &  0 \\
 0 &   0 & -1 &  0 &  0 &  0 \\
 0 &   0 &  0 & -1 &  0 &  0 \\
 0 &   1 &  0 &  0 &  0 &  0 \\
 1 &   0 &  0 &  0 &  0 &  0 
\end{pmatrix*}\, .
\end{equation}
The bosonic case is similar except there are no minus signs in the Unitaries so they are simply permutation matrices.

\section{ Invariant subspaces of non-universal puzzles.}\label{app:invariant}

{One consequence of the square root SWAPs being subuniversal on $\text{SU}(6)$ is the state space is divided into several disconnected subspaces.  As an example, consider the action of arbitrary $z$ rotations on a qubit. This is subuniversal and results in a set of invariant subspaces parametrized by the polar angle on the Bloch sphere. If a set of operations is universal, then the state space must be connected, and thus there is only one subspace. 

We define the solving subspace to be all states reachable by square root SWAPs applied to \solved. If instead we started from a different state that is not in the solving subspace, the puzzle could never be returned to the solved state by square root SWAPs. This is analogous to peeling off the stickers of a Rubik's cube and placing them randomly. Although it is possible that it may result in a legal scramble, it is most likely that the puzzle will be unsolvable. 

 For the bosonic puzzle there is a single invariant state, the equal superposition state,
\begin{equation}
    |\Psi\rangle = \frac{1}{\sqrt{6}}\left(\zero{15pt}+\one{15pt}+\two{15pt}+\dots+\five{15pt}\right).
\end{equation}
Any unitary operation that is not invariant on this state cannot be generated by the square root SWAPs. For the fermionic puzzle the equal superposition state is not completely invariant, but it does generate a subspace that is separate from the solving subspace. This subspace is characterized by states that are invariant under color SWAP as defined in \cref{app:centralizer}
}

\section{Set of $6\times6$ matrices that commute with all square root SWAPs}\label{app:centralizer}
Consider the Fermionic version of the puzzle. We can explicitly construct a set of matrices that commute with every element of the puzzle's group $\{\mathbb{H}_k\}$. Note that it is sufficient to show that a matrix commutes with each individual square root SWAP because it then must also commute with all products of square root SWAPs and thus the entire group. It is not to hard to show that the set of all matrices that commute with all four square root SWAPs must be of the form
\begin{equation}
   \mathbb{K}(a,b,c) =  \begin{pmatrix*}[r]  a & b & -c & -c & \phantom{-}c & \phantom{-}c \\ b & a & -c & -c & c & c \\ -c & -c & a & b & c & c \\ -c & -c& b & a & c & c \\ c & c & c & c & a & b \\ c & c & c & c & b & a  \end{pmatrix*}\,,
\end{equation}
where $a,b,c\in \mathbb{C}$. This defines a three-dimensional subspace of $6\times6$ matrices. We see that $\mathbb{K}(1,0,0)$ is the identity matrix, which trivially commutes with every matrix.
If we instead only pick out the $b$ elements we get 
\begin{equation}
\mathbb{K}(0,1,0) = \begin{pmatrix}
 0 &   1 &  0 &  0 &  0 &  0 \\
 1 &   0 &  0 &  0 &  0 &  0 \\
 0 &   0 &  0 &  1 &  0 &  0 \\
 0 &   0 &  1 &  0 &  0 &  0 \\
 0 &   0 &  0 &  0 &  0 &  1 \\
 0 &   0 &  0 &  0 &  1 &  0 
\end{pmatrix}\, .
\end{equation}
This matrix is a permutation matrix which squares to the identity. It turns out that this operation swaps the colors of the particles in the puzzle. For that reason we will call this the color SWAP.

Finally, the last linearly independent matrix that commutes with the square root swaps is $\mathbb{K}(0,0,1)$. This matrix is hermitian, but not unitary and thus is not a legal operation on the puzzle.

\section{Fermionic vs Bosonic SWAPs}\label{app:vacuumswaps}

Lets imagine a the simplest case where we would have a swap of identical particles. We consider 2 sites each with at most 1 particle. For now consider only one species.  We can write down distinguishable state as 
\begin{equation}
    |00\rangle,~|0,1\rangle,~|1,0\rangle,~|1,1\rangle
\end{equation}
where the first indecies account for particle location and the numbers indicate the number of paricles in that location, i.e $|0,1\rangle$ means no particle in the first location and one particle in the second location. 

For bosons the Full swap matrix acting on this Hilbert space is a permutation matrix with explicit form,
\begin{equation}
    \mathbb{S}_b = \begin{pmatrix}
        1&0&0&0\\
        0&0&1&0\\
        0&1&0&0\\
        0&0&0&1
    \end{pmatrix}
\end{equation}
which acts on the states as follows 
\begin{equation}
\begin{aligned}
     \mathbb{S}_b|00\rangle &\to |00\rangle\\
     \mathbb{S}_b|01\rangle &\to |10\rangle\\
     \mathbb{S}_b|10\rangle &\to |01\rangle\\
     \mathbb{S}_b|11\rangle &\to |11\rangle\,.
\end{aligned}
\end{equation}
The only difference with fermions is in the last element 
\begin{equation}
    \mathbb{S}_f = \begin{pmatrix}
        1&0&0&0\\
        0&0&1&0\\
        0&1&0&0\\
        0&0&0&-1
    \end{pmatrix}\,.
\end{equation}
This shows that swapping two instances of fermionic vacuum does not pick up a phase, i.e $\mathbb{S}_f|00\rangle \to |00\rangle$. Additionally swapping fermions with vacuum picks up no phase, but swapping two identical fermions picks up a minus sign. 

Now if we go back to our 2x2 puzzle with two green particles and two blue particles we will quickly rederive the swapping relations. First we must note that swapping a green and blue fermions is really an emergent action. This action is a result of  swapping a green particle in location 1 with green  vacuum in location 2 and a blue particle in location 2 with blue vacuum in location 1. Each swap of particle and vacuum with pick up no phase since for both bosons and fermions $\mathbb{S}|01\rangle \to|10\rangle$.  

So for both fermions and bosons it is true that 
\begin{subequations}
\begin{align}
\mathbb{S}_U\two{15pt} &= +1\five{15pt} \\
\mathbb{S}_U\three{15pt} &= +1\four{15pt} \\
\mathbb{S}_U\four{15pt} &= +1\three{15pt} \\
\mathbb{S}_U\five{15pt} &= +1\two{15pt} \,.
\end{align}
\end{subequations}
But they differ when identical particles are exchanged, e.g for bosons
\begin{subequations}
\begin{align}
\mathbb{S}_U\zero{15pt} &= +1\zero{15pt} \\
\mathbb{S}_U\one{15pt} &= +1\one{15pt}, 
\end{align}
\end{subequations}
and for fermions
\begin{subequations}
\begin{align}
\mathbb{S}_U\zero{15pt} &= -1\zero{15pt} \\
\mathbb{S}_U\one{15pt} &= -1\one{15pt}.
\end{align}
\end{subequations}

Now a final question is what happens in the situation where we have green fermions and blue bosons. It is easy to show that all the situations in which no identical particles are swapped results in no phase since this is true for both bosons and fermions. Now for a specific swap we just need to determine which particles are swapped. For example if we swap the top particles of $\solved$ we would be swapping green particles which are fermions so $\mathbb{S}_U\solved = -\solved$. But if we apply the same swap to $\one{15pt}$ we would swap the blue particles which are bosons so no phase would be introduced. 
Similarly if we remove the bosons and are left with two empty sites of fermionic vacuum we would have the same behavior. This follows directly from our previous note that fermionic vacuum can be exchanged without picking up a phase.

\section{General puzzles: universality and color permutations}\label{app:universal}

\subsection{\nbym{n}{1} puzzles - universality}

We can embed our puzzle into a family of puzzles. For simplicity, we consider \nbym{n}{1} family of bosonic puzzles with one green particle and $n-1$ blue particles. Such puzzles have Hilbert space dimension $n$ and basis states $\ket{i}$ for $i\in [0,n-1]$. For actions on the puzzle we allow: nearest neighbor SWAPs 
\begin{equation}
\mathbb{S}_{i,i+1}= \op{i}{i+1}+\op{i+1}{i} + \sum_{k\in \mathcal{N}/\{i,i+1\}}\op{k}{k}
\end{equation}
where $i\in \{0,1,\ldots n-2 \}$ and $\mathcal{N}\coloneqq \{0,1,2,\ldots, n-1\}$. In total there are $n-1$ swaps which produce $n-1$ square root SWAPs $\mathbb{H}_{i} = \exp[i(\pi/4) \mathbb{S}_{i,i+1}]$. We also construct a diagonal gate with one $-1$ i.e. $\mathbb{P}_N = {\rm diag}(1,1,\ldots, -1)$.

First, we start with an anomaly in the puzzle family. The \nbym{2}{1} puzzle has a 2-dimensional Hilbert space, easily mapped onto a qubit.  It only has one allowed square root SWAP,  $\mathbb{H}_{01}$, analogous to a rotation about $X$ by $\pi/2$. The universality check with $\mathbb{H}_{01}$ (a rotation about $X$ by $\pi/2$) and $\mathbb{P}_2$ (the Pauli $Z$ gate) fails as $X_{\pi/2}$ and $Z$ are not universal for a qubit. If we include the $T$ gate instead of $\mathbb{P}_2$ the puzzle is universal. This appears to be the only puzzle in this family where $\mathbb{P}_N$ alone does not ensure universality.

We numerically ran the universality test for $n\in \{3,4,\ldots,9\}$ and it shows that the gate set $G =\{\mathbb{H}_{i}\}\cup \mathbb{P}_n$ is universal. Based on this we conjecture that the gate set $G$ is universal for any \nbym{n}{1} with 1 green and $n-1$ blue particles, for $n>2$. 

We can formulate a similar conjecture for \nbym{n}{m} puzzles from the following observation. The \nbym{n}{1} puzzles can be reshaped into \nbym{n'}{m'} puzzles where $n = n'\times m'$.
Based on this, we conjecture that for any \nbym{n}{m} puzzle with at least one repeated color, the gate set consisting of all nearest neighbor square root SWAPs and $\mathbb{P}_n$ is universal.

\subsection{\nbym{2}{3} puzzle color permutations}
A \nbym{2}{3} puzzle with $3$ green and $3$ blue particles can be mapped to a qudit with a 20-dimensional Hilbert space. We find that there is a unitary matrix that commutes with every element of the puzzle's group $\{\mathbb{H}_k\}$ that is analogous to the color SWAP in \cref{app:centralizer}. Essentially it SWAPs all the green and blue particles.

Next consider a \nbym{2}{3} puzzle with $2$ green, $2$ blue, and $2$  red particles. This can be mapped to a qudit with a 90-dimensional Hilbert space. We find that there are 5 unitary matrices that commute with every element of the puzzle's group $\{\mathbb{H}_k\}$. There are three two-color permutations and two three-color permutations. These symmetries and more will clearly exist in larger puzzles.

\section{Solving advantage in certain puzzles}\label{app:quant_adv}

Here, we present a small numerical study on solving advantages.  A key challenge is that computational time grows with the number of moves and board dimensions, limiting exploration of larger \nbym{n}{m} boards. Because of this limitation, we focus on two different classes of puzzles. First we examine \nbym{n}{1} puzzles for $n \in \{2, 3, 4, 5\}$ with one green particle and $n-1$ blue particles. We also examine a class of \nbym{2}{2} puzzles with increasing Hilbert space dimension, which is acheived by changing the number of colored particles in the puzzle.

\subsection{\nbym{n}{1} puzzles - solving advantage}

\begin{figure}[!th]
    \centering
    \includegraphics[width=0.9\columnwidth]{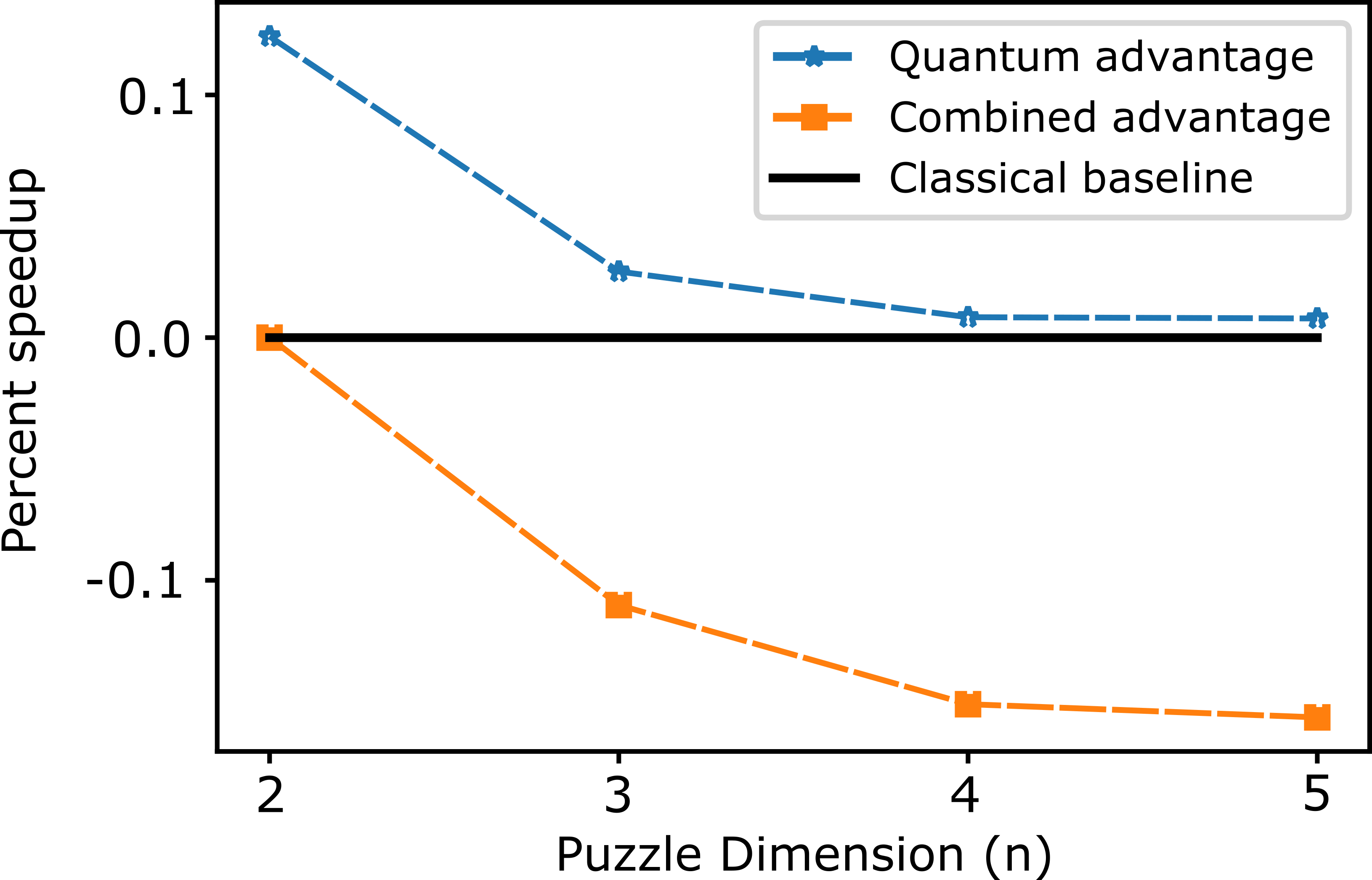}
    \caption{ Comparison of quantum and combined solvers relative to the classical solver. We compute optimal solutions for $n \times 1$ puzzles with $n \in \{2,3,4,5\}$ using all three solvers. The graph displays the percent difference in average move counts: the black line represents the classical strategy baseline, the blue line shows the quantum solver versus the classical, and the orange line shows the combined solver versus the classical. A positive percentage indicates that the classical solver performed better, while a negative percentage indicates that the quantum or combined solver was superior. Each data point is averaged over 1350 scrambles. For smaller puzzles, the classical solver outperforms the quantum solver, but this gap nearly disappears for the $5 \times 1$ puzzle. The combined solver consistently outperforms the classical solver, except at $n=2$, where the classical strategy is optimal and the combined solver merely matches its performance. 
    Due to numerical constraints we did not simulate for $n>5$.}

    \label{fig:dim_advant}
\end{figure}

Consider the \nbym{n}{1} puzzles for $n \in \{2, 3, 4, 5 \}$ with one green particle and $n-1$ blue particles. These puzzles are the most classical and simplest puzzles with $n-1$ SWAPs and $n$ basis states. In \cref{fig:dim_advant}, we show for the \nbym{n}{1} puzzles increasing Hilbert space dimension corresponds to an increasing advantage for the combined solvers compared to the classical solver. This is presented as the percentage difference in the average move count between the combined solver {\bf relative} to the classical solvers move count for various values of $n$. The quantum solver was not able to have an advantage over the classical solver for any of the \nbym{n}{1} puzzles we were able to simulate. This may be because the \nbym{n}{1} puzzles require relatively long classical solutions, of order $n$, but do not have many states to spread amplitude over. The long solutions penalize the slower moves of the quantum solver, and the ability to concentrate amplitude is not able to offset this. 

\subsection{\nbym{2}{2} puzzles - solving advantage}

\begin{figure}[!th]
    \centering
    \includegraphics[width=0.9\columnwidth]{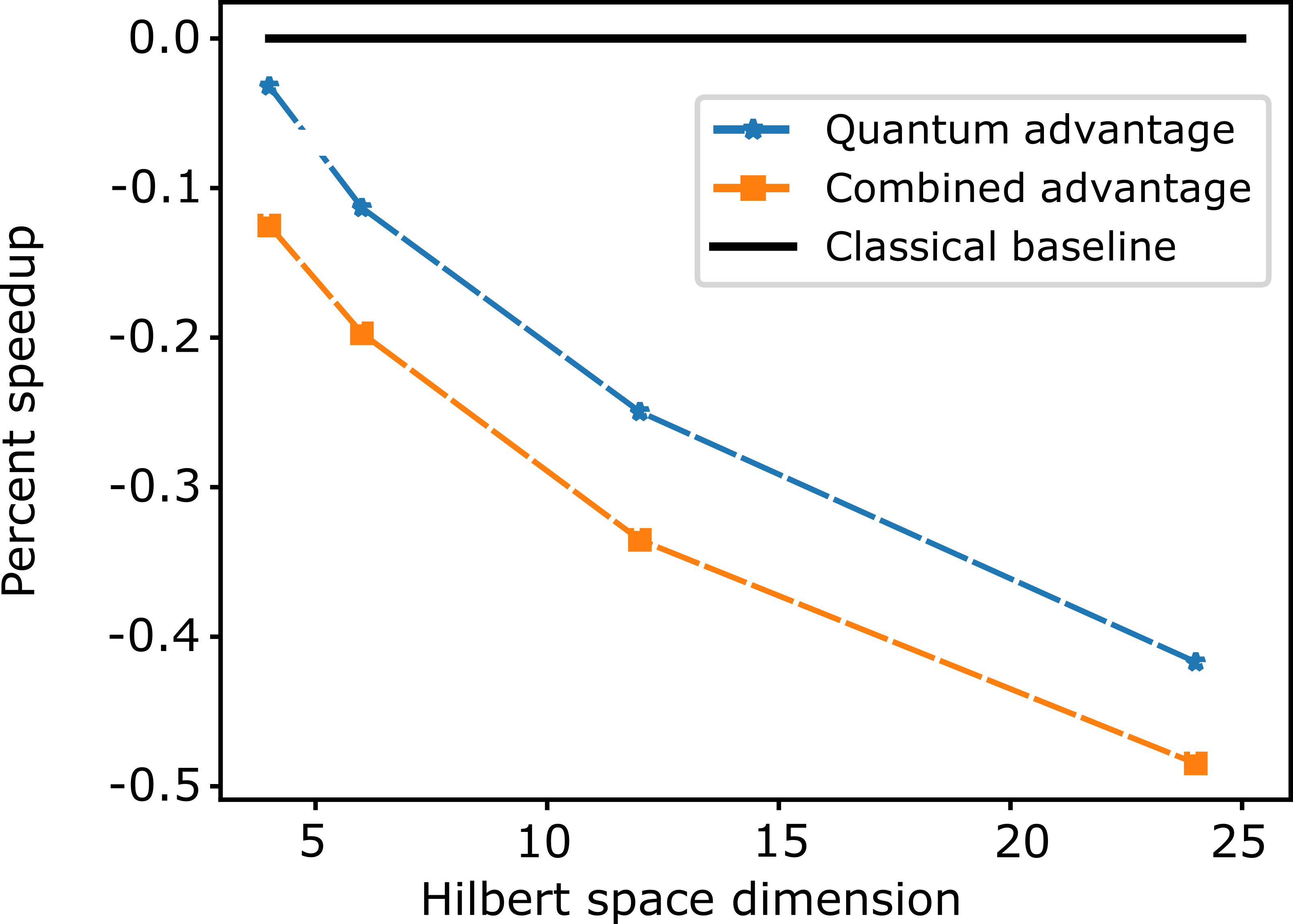}
    \caption{Percent difference between quantum and classical solver (blue) alongside the percent difference between combined and classical solver (orange). A negative percentage indicates a quantum speedup. For \nbym{2}{2} puzzles with increase Hilbert space dimension we see an increase in the advantage of both the quantum and combined solver.  Here we see a large quantum advantage of  $\sim 40\%$ for the largest \nbym{2}{2} puzzle. Each data point is averaged over 500 scrambles.   
    }\label{fig:dim_advant_2}
\end{figure}

 For the \nbym{2}{2} puzzles the Hilbert space dimension is increased by changing the colors of the four particles in the puzzle. The set of possible sets of four particles is: 3 green 1 blue, 2 green 2 blue, 2 green 1 blue 1 yellow, and 1 green 1 blue 1 yellow 1 red. These have Hilbert space dimensions 4, 6, 12, and 24 respectively. By keeping the SWAPs and puzzle geometry fixed we can directly see how Hilbert space dimension affects solving. In \cref{fig:dim_advant_2} we plot the percent speed-up of the quantum and combined solver for the four puzzles we mention here. Not only do we see the quantum speedup increase with Hilbert space dimension, but we observe a quite large quantum advantage of roughly 40\% for the largest Hilbert space. This supports our conjecture that when SWAPs are fixed, increasing Hilbert space dimension results in an increasing quantum advantage.

\section{The 3D Treatment of the cube}\label{app:3D_cube_rotations}

Our goal is to show that rotating a collection of particles simply permutes the board states and thus justifies the permutation treatment given in the main text. For simplicity, we will consider the particles in this section to be Bosons.

Let's start by defining the location of the quantum particles in 3D by the position vector. Because we will need to specify 16 locations we introduce a subscript e.g. a particle at location 1 is at the position 
\begin{equation}
    \mathbf{r}_1 = \begin{pmatrix}
        x_1\\ y_1\\ z_1
    \end{pmatrix}\, ,
\end{equation}
these locations are fixed and don't depend on the color of the particle, see \cref{fig:cube_location_labels}.

\begin{figure}[ht!]
\raisebox{4em}{
\includestandalone[width=0.2\columnwidth]{axis}} \hspace{5pt}
\includestandalone[width=0.33\columnwidth]{cube_2x2x1_labels} \hspace{15pt}
\includestandalone[width=0.27\columnwidth]{b_lab} 
\caption{The co-ordinate system. The co-ordinates remain fixed, but the color of particles may change at a given location. For puzzle state $\ket{0}$ the particles at location $r_i$ are: red for $i\in \{ 1,2\}$, yellow for $i\in\{3,4,5,6\}$, orange for $i\in\{7,8\}$, white for $i\in\{9,10,11,12\}$, green for $i\in\{13,14\}$, 
 and blue for $i\in\{15,16\}$.}\label{fig:cube_location_labels}
\end{figure}

We limit rotations to $\pm \pi$ along any axis to prevent changing the puzzle's shape, keeping it within the six board configurations. Now we want to determine how a rotation, e.g. rotation about the $z$-axis acts on the collection of particles. Recall that the angular momentum operator, in this case
\begin{equation}
 \hat L_z = \hat x \hat p_y - \hat y \hat p_x  \, , 
\end{equation}
is the generator of rotations in the space of physical states. The corresponding rotation operator is $R_z(\theta) = \exp(- i \theta \hat L_z)$, which can be written in terms of its SO(3) action
\begin{equation}
R_z(\theta)  = 
\begin{pmatrix}
\cos\theta & -\sin\theta & 0\\
\sin\theta & \cos\theta & 0\\
0          &  0         & 1
\end{pmatrix}
\end{equation}
so that 
\begin{equation}\label{eq:wfn_rot_app}
R_z(\theta) \psi(\mathbf{r}_i) = 
\psi\begin{pmatrix}
    x_i \cos\theta  -y_i \sin\theta \\  x_i\sin\theta + y_i\cos\theta \\ z_i
\end{pmatrix} = \psi(\mathbf{r}_i')\, ,
\end{equation}
where $\psi(\mathbf{r}_i)$ is the position wavefunction of the $i$'th particle.

However our moves $U,R$ are not simply a rotation about the $z$ axis. The move $U$ for example, is a rotation about the $z$ axis restricted to any positive $z$ coordinate, i.e. $z>0$, so that the cubes below $z=0$ don't move. We denote these restricted rotations as  
\begin{equation}
\hat U = R_z^U(\theta) \,,
\end{equation}
and the related rotation about the positive $y$ axis would be $\hat R = R_y^R(\theta)$.

Now consider the quantum state of the bosons. To be explicit we consider point like bosons denote the wavefunction of a single particle at location $\mathbf{r}_1$ is denoted by $\psi^{\rm color}_{\rm label}({\rm location})$. So if the third yellow particle was at location $\mathbf{r}_1$
\begin{equation}
    \psi_{3}^{y}(\mathbf{r}_1) \, ,
\end{equation}
where $\int d^3\mathbf{r} |\psi_{3}^{y}(\mathbf{r}_1)|^2  =1$.
To simplify the description of indistinguishable particles we will denote the wavefunction of four e.g. yellow particles at four locations
\begin{equation}
    \psi^y(3,4,5,6) = \psi_{1}^{y}(\mathbf{r}_3) \psi_{2}^{y}(\mathbf{r}_4) \psi_{3}^{y}(\mathbf{r}_5) \psi_{4}^{y}(\mathbf{r}_6) \,.
\end{equation}
Using this notation we can denote the (anti) symmetrized wavefunction as
\begin{equation}
\begin{aligned}
    \Psi^y(3,4,5,6) \propto & \,\, \psi^y(3,4,5,6) + \psi^y(4,3,5,6) +\\
     & \,\, \psi^y(3,4,6,5) + \psi^y(4,3,6,5)\,.
\end{aligned}
\end{equation}
The solved state wavefunction will be denoted as
\begin{equation}
\begin{aligned}
    \Phi_0 = &\, \Psi^r(1,2)\Psi^y(3,4,5,6) \Psi^o(7,8)\times \\
    &\,\Psi^w(9,10,11,12) \Psi^g(13,14) \Psi^b(15,16) \, ,
\end{aligned}
\end{equation}
while for the first state it would be
\begin{equation}
\begin{aligned}
    \Phi_1 = &\, \Psi^r(1,2)\Psi^y(5,6,11,12) \Psi^o(7,8)\times \\
    &\,\Psi^w(3,4,9,10) \Psi^g(14,15) \Psi^b(13,16) \, ,
\end{aligned}
\end{equation}
and similarly $\Phi_i$ for the $i$'th board state.

Defining the correctly (anti) symmetrized wavefunction for the 16 particles and showing that it transforms correctly is tedious, so we present those calculations in \cref{app:explicit_3dcube} for the following case
\begin{align}
\Phi_1 = R^U_z(\pm \pi) \Phi_0 .
\end{align}
With some more work, one can show that
\begin{equation}
\begin{aligned}
    \Phi_2 &= R_y^R(\pi) \Phi_0\\
    \Phi_3 &= R_y^R(\pi) R_z^U(\pi) \Phi_0\\
    \Phi_4 &= R_z^U(\pi) R_y^R(\pi) \Phi_0\\
    \Phi_5 &=  R_z^U(\pi)R_y^R(\pi) R_z^U(\pi) \Phi_0 ,
\end{aligned}
\end{equation}
which is consistent with the classical moves \cref{eq:class_moves} and the permutation description. That means we have a fully quantum description, in 3D, of our puzzle.

In order to add an analogous move to the \halfswap\ we need to be able to implement the operation
\begin{equation}
  \mathbb{R}_z^U  = \exp \big [i \tfrac{\pi}{2} R_z^U(\pi)/2\big ] =  \cos\big ( \tfrac{\pi}{4}\big) I + i \sin\big ( \tfrac{\pi}{4}\big) R_z^U(\pi).
\end{equation}
Physically implementing such a rotation in 3D seems hard. One possible way is to do a controlled rotation gate, controlled of e.g. a qubit in the plus state, and measure in the qubit $x$ basis. Then conditional on getting the $+$ outcome the \halfswap\ operation would be enacted.

\subsection{`U' move on the solved state}\label{app:explicit_3dcube}
In this subsection we show explicitly that
\begin{equation*}
 \Phi_1 = R^U_z(\pm \pi) \Phi_0 .   
\end{equation*}

\subsubsection{Red and Orange particles}
From looking at \cref{fig:cube_location_labels} we see that the wavefunction of the red particles in the solved state is
\begin{equation}
 \Psi^r(1,2) \propto   \psi^r(1,2) + \psi^r(2,1) . 
\end{equation}
We can see in \cref{fig:cube_states} that the red particles remain in the same location after the rotation so lets check that. Assume that 
\begin{equation}
    \mathbf{r}_1 = \begin{pmatrix}
        -x_1\\ 0\\ z_1
    \end{pmatrix} , \,    \mathbf{r}_2 = \begin{pmatrix}
        x_1\\ 0\\ z_1
    \end{pmatrix}\, ,
\end{equation}
then using \cref{eq:wfn_rot_app} we can show 
\begin{equation}
  \Psi^r(1,2) = R_z^U(\pm\pi) \Psi^r(1,2) \propto   \psi^r(1,2) + \psi^r(2,1) , 
\end{equation}
which just means that there is no change, as expected. The orange particles are completely unaffected by this rotation, so we don't consider them.

\subsubsection{Green and Blue particles} 
For the green and blue particles, we define locations
\begin{equation}
\begin{aligned}
\mathbf{r}_{13} = &\begin{pmatrix}
        0 \\ -y_{13} \\ z_{13}
    \end{pmatrix}, \, \, \, 
    \mathbf{r}_{14} = \begin{pmatrix}
        0\\ -y_{13}\\ -z_{13}
    \end{pmatrix},\\ 
\mathbf{r}_{15} = &\begin{pmatrix}
        0 \\ \phantom{-}y_{13} \\ z_{13}
    \end{pmatrix}, \, \, \, 
    \mathbf{r}_{16} = \begin{pmatrix}
        0\\ y_{13}\\ -z_{13}
    \end{pmatrix}. 
\end{aligned}
\end{equation}
The wavefunctions before the rotation are
\begin{equation}
\begin{aligned}
\Psi^g(13,14) & \propto   \psi^g(13,14) + \psi^g(14,13) , \\
\Psi^b(15,16) & \propto   \psi^b(15,16) + \psi^b(16,15) , 
\end{aligned}
\end{equation}
and after
\begin{equation}
\begin{aligned}
\Psi^g(14,15) &= R_z^U(\pm\pi)\Psi^g(13,14) \\ 
\Psi^b(13,16) &= R_z^U(\pm\pi)\Psi^b(15,16)  \, ,
\end{aligned}
\end{equation}
which is consistent with the $\ket{1}$ board state.

\subsubsection{White and Yellow particles}
Initally the yellow particles are at the locations
\begin{equation}
\begin{aligned}
\mathbf{r}_{3} = &\begin{pmatrix}
        x_{3} \\ -y_{3} \\ z_{3}
    \end{pmatrix}, \, \, \, 
    \mathbf{r}_{4} = \begin{pmatrix}
        x_{3} \\ y_{3}\\ z_{3}
    \end{pmatrix},\\ 
\mathbf{r}_{5} = &\begin{pmatrix}
        x_{3} \\ -y_{3} \\ -z_{3}
    \end{pmatrix}, \, \, \, 
    \mathbf{r}_{6} = \begin{pmatrix}
        x_{3} \\ y_{13}\\ -z_{3}
    \end{pmatrix}, 
\end{aligned}
\end{equation}
and the white particles are at
\begin{equation}
\begin{aligned}
\mathbf{r}_{11} = &\begin{pmatrix}
        -x_{3} \\ -y_{3} \\ z_{3}
    \end{pmatrix}, \, \, \, 
    \mathbf{r}_{12} = \begin{pmatrix}
        -x_{3} \\ y_{3}\\ z_{3}
    \end{pmatrix},\\ 
\mathbf{r}_{9} = &\begin{pmatrix}
        -x_{3} \\ -y_{3} \\ -z_{3}
    \end{pmatrix}, \, \, \, 
    \mathbf{r}_{10} = \begin{pmatrix}
        -x_{3} \\ y_{13}\\ -z_{3}
    \end{pmatrix}. 
\end{aligned}
\end{equation}
The wave function of the yellow particles in the solved state is
\begin{equation}
\begin{aligned}
    \Psi^y(3,4,5,6) \propto & \,\, \psi^y(3,4,5,6) + \psi^y(4,3,5,6) +\\
     & \,\, \psi^y(3,4,6,5) + \psi^y(4,3,6,5)\,,
\end{aligned}
\end{equation}
as $R_z^U$ only acts on locations 3 and 4 we only expect those particles to transform and similarly for the white particles. After acting the rotation we find
\begin{equation}
\begin{aligned}
\Psi^y(5,6,11,12) &= R_z^U(\pm\pi)\Psi^y(3,4,5,6) \\ 
\Psi^w(3,4,9,10) & = R_z^U(\pm\pi)\Psi^w(9,10,11,12)  \, .
\end{aligned}
\end{equation}

 In summary, what we have shown is that 
\begin{align}
\Phi_1 = R^t_z(\pm \pi) \Phi_0 .
\end{align}

\section{Cube Qudit mapping}\label{apx:cube_qudit}

As before we map the six cube states to a 6 dimensional qudit. The board states are represented by the following column vectors
\begin{align*}
\ket{0} &= \begin{pmatrix}
 1 & 0 & 0 & 0 & 0 & 0 
\end{pmatrix}^T\\
\ket{1} &= \begin{pmatrix}
 0 & 1 & 0 & 0 & 0 & 0 
\end{pmatrix}^T\\
\ket{2} &= \begin{pmatrix}
 0 & 0 & 1 & 0 & 0 & 0 
\end{pmatrix}^T\\
\ket{3} &= \begin{pmatrix}
 0 & 0 & 0 & 1 & 0 & 0 
\end{pmatrix}^T\\
\ket{4} &= \begin{pmatrix}
 0 & 0 & 0 & 0 & 1 & 0 
\end{pmatrix}^T\\
\ket{5} &=\begin{pmatrix}
 0 & 0 & 0 & 0 & 0 & 1 
\end{pmatrix}^T .
\end{align*}
To determine the matrix elements of $P_U$ we the action $U$ on the cube and map its actions on the basis states and find
\begin{equation}\label{eq:cube_perm_U}
\begin{aligned}
\mathbb{P}_U \ket{0} &= \ket{1} \\
\mathbb{P}_U \ket{1} &= \ket{0} \\
\mathbb{P}_U \ket{2} &= \ket{4} \\
\mathbb{P}_U \ket{3} &= \ket{5} \\
\mathbb{P}_U \ket{4} &= \ket{2} \\
\mathbb{P}_U \ket{5} &= \ket{3} \,.
\end{aligned}
\end{equation}
It's similarly straightforward to show that
\begin{equation}\label{eq:cube_perm_R}
\begin{aligned}
\mathbb{P}_R \ket{0} &= \ket{2} \\
\mathbb{P}_R \ket{1} &= \ket{3} \\
\mathbb{P}_R \ket{2} &= \ket{0} \\
\mathbb{P}_R \ket{3} &= \ket{1} \\
\mathbb{P}_R \ket{4} &= \ket{5} \\
\mathbb{P}_R \ket{5} &= \ket{4} \,.
\end{aligned}
\end{equation}
From these relations we define the permutation matrix (assuming bosonic statistics)
\begin{equation}
\mathbb{P}_U = \begin{pmatrix}
 0 &   1 &  0 &  0 &  0 &  0 \\
 1 &   0 &  0 &  0 &  0 &  0 \\
 0 &   0 &  0 &  0 &  1 &  0 \\
 0 &   0 &  0 &  0 &  0 &  1 \\
 0 &   0 &  1 &  0 &  0 &  0 \\
 0 &   0 &  0 &  1 &  0 &  0 
\end{pmatrix}\, ,
\end{equation}
in this case the inverse $(\mathbb{P}_U )^{-1} = \mathbb{P}_U^T = \mathbb{P}_U$.
The permutation matrix for the $\mathbb{P}_R$ action is 
\begin{equation}
\mathbb{P}_R = \begin{pmatrix}
 0 &   0 &  1 &  0 &  0 &  0 \\
 0 &   0 &  0 &  1 &  0 &  0 \\
 1 &   0 &  0 &  0 &  0 &  0 \\
 0 &   1 &  0 &  0 &  0 &  0 \\
 0 &   0 &  0 &  0 &  0 &  1 \\
 0 &   0 &  0 &  0 &  1 &  0 
\end{pmatrix}\, .
\end{equation}

\end{document}

%% file: cube_solved.tex
    \RubikFaceUp
        {rc}{rc}{rc}
        {rc}{rc}{rc}
        {rc}{rc}{rc}
    \RubikFaceRight
        {bc}{bc}{bc}
        {bc}{bc}{bc}
        {bc}{bc}{bc}
    \RubikFaceFront
        {yc}{yc}{yc}
        {yc}{yc}{yc}
        {yc}{yc}{yc}
   \ShowCube{3.5cm}{0.7}{\DrawRubikCube}

%% file: cube_scram.tex
    \RubikFaceUp
        {bc}{oc}{oc}
        {gc}{wc}{gc}
        {gc}{oc}{oc}
    \RubikFaceRight
        {gc}{rc}{bc}
        {gc}{rc}{yc}
        {yc}{bc}{oc}
    \RubikFaceFront
        {wc}{wc}{yc}
        {yc}{gc}{wc}
        {bc}{rc}{rc}
   \ShowCube{3.5cm}{0.7}{\DrawRubikCube}

%% file: slide_solved.tex
\begin{tikzpicture}[every node/.style={minimum size=1cm-\pgflinewidth, outer sep=0pt}]
    \node[fill=kc] at (0.5,+2.5) {};
    \node[fill=gc] at (1.5,+2.5) {};
    \node[fill=gc] at (2.5,+2.5) {};
    \node[fill=bc] at (0.5,+1.5) {};
    \node[fill=bc] at (1.5,+1.5) {};
    \node[fill=bc] at (2.5,+1.5) {};
    \node[fill=yc] at (0.5,+0.5) {};
    \node[fill=yc] at (1.5,+0.5) {};
    \node[fill=yc] at (2.5,+0.5) {};
    \draw[step=1cm,color=black,line width = 0.7mm] (-0.025,-0.025) grid (3.025,3.025);
\end{tikzpicture}

%% file: slide_scram.tex
\begin{tikzpicture}[every node/.style={minimum size=1cm-\pgflinewidth, outer sep=0pt}]
    \node[fill=gc] at (0.5,+2.5) {};
    \node[fill=bc] at (1.5,+2.5) {};
    \node[fill=gc] at (2.5,+2.5) {};
    \node[fill=bc] at (0.5,+1.5) {};
    \node[fill=yc] at (1.5,+1.5) {};
    \node[fill=bc] at (2.5,+1.5) {};
    \node[fill=yc] at (0.5,+0.5) {};
    \node[fill=kc] at (1.5,+0.5) {};
    \node[fill=yc] at (2.5,+0.5) {};
    \draw[step=1cm,color=black,line width = 0.7mm] (-0.025,-0.025) grid (3.025,3.025);

\end{tikzpicture}

%% file: slide_quant.tex
$\Bigg | \raisebox{-1.15em}{\resizebox{0.98cm}{!}{
\begin{tikzpicture}[every node/.style={minimum size=1cm-\pgflinewidth, outer sep=0pt}]
    \node[fill=kc] at (0.5,+2.5) {};
    \node[fill=gc] at (1.5,+2.5) {};
    \node[fill=gc] at (2.5,+2.5) {};
    \node[fill=bc] at (0.5,+1.5) {};
    \node[fill=bc] at (1.5,+1.5) {};
    \node[fill=bc] at (2.5,+1.5) {};
    \node[fill=yc] at (0.5,+0.5) {};
    \node[fill=yc] at (1.5,+0.5) {};
    \node[fill=yc] at (2.5,+0.5) {};
    \draw[step=1cm,color=black,line width = 0.5mm] (-0.025,-0.025) grid (3.025,3.025);
\end{tikzpicture}
}}\Bigg \rangle$

%% file: slide_super.tex
$\propto$
$\Bigg | \raisebox{-1.15em}{\resizebox{0.98cm}{!}{
\begin{tikzpicture}[every node/.style={minimum size=1cm-\pgflinewidth, outer sep=0pt}]
    \node[fill=kc] at (0.5,+2.5) {};
    \node[fill=gc] at (1.5,+2.5) {};
    \node[fill=gc] at (2.5,+2.5) {};
    \node[fill=bc] at (0.5,+1.5) {};
    \node[fill=bc] at (1.5,+1.5) {};
    \node[fill=bc] at (2.5,+1.5) {};
    \node[fill=yc] at (0.5,+0.5) {};
    \node[fill=yc] at (1.5,+0.5) {};
    \node[fill=yc] at (2.5,+0.5) {};
    \draw[step=1cm,color=black,line width = 0.5mm] (-0.025,-0.025) grid (3.025,3.025);
\end{tikzpicture}
}}\Bigg \rangle$
$+$\hspace{2pt}
$\Bigg | \raisebox{-1.15em}{\resizebox{0.98cm}{!}{
\begin{tikzpicture}[every node/.style={minimum size=1cm-\pgflinewidth, outer sep=0pt}]
    \node[fill=gc] at (0.5,+2.5) {};
    \node[fill=bc] at (1.5,+2.5) {};
    \node[fill=gc] at (2.5,+2.5) {};
    \node[fill=bc] at (0.5,+1.5) {};
    \node[fill=yc] at (1.5,+1.5) {};
    \node[fill=bc] at (2.5,+1.5) {};
    \node[fill=yc] at (0.5,+0.5) {};
    \node[fill=kc] at (1.5,+0.5) {};
    \node[fill=yc] at (2.5,+0.5) {};
    \draw[step=1cm,color=black,line width = 0.5mm] (-0.025,-0.025) grid (3.025,3.025);
\end{tikzpicture}
}}\Bigg \rangle $

%% file: cube_2x2x1.tex
\pgfmathsetmacro\radius{0.1}%

\def\sidecolor{bc}%
\def\frontcolor{yc}%
\def\topcolor{rc}%
\foreach \myPsi in {130}{%
    \tdplotsetmaincoords{70}{\myPsi}%
    \begin{tikzpicture}[line join=round]%
        \begin{scope}[tdplot_main_coords]%
            \filldraw [canvas is yz plane at x=1.5] (-0.5,-0.5) rectangle (1.5,1.5);%
            \filldraw [canvas is xz plane at y=1.5] (0.5,-0.5) rectangle (1.5,1.5);%
            \filldraw [canvas is yx plane at z=1.5] (-0.5,0.5) rectangle (1.5,1.5);%
            \foreach \X in {-0.5,0.5}{%
                \foreach \Y in {-0.5,0.5}{%
                   \draw [canvas is yz plane at x=1.5,shift={(\X,\Y)},fill=\frontcolor] (0.5,0) -- ({1-\radius},0) arc (-90:0:\radius) -- (1,{1-\radius}) arc (0:90:\radius) -- (\radius,1) arc (90:180:\radius) -- (0,\radius) arc (180:270:\radius) -- cycle;}}
            \draw [canvas is xz plane at y=1.5,shift={(0.5,-0.5)},fill=\sidecolor] (0.5,0) -- ({1-\radius},0) arc (-90:0:\radius) -- (1,{1-\radius}) arc (0:90:\radius) -- (\radius,1) arc (90:180:\radius) -- (0,\radius) arc (180:270:\radius) -- cycle;%
            \draw [canvas is xz plane at y=1.5,shift={(0.5,0.5)},fill=\sidecolor] (0.5,0) -- ({1-\radius},0) arc (-90:0:\radius) -- (1,{1-\radius}) arc (0:90:\radius) -- (\radius,1) arc (90:180:\radius) -- (0,\radius) arc (180:270:\radius) -- cycle;%
            \draw [canvas is yx plane at z=1.5,shift={(0.5,0.5)},fill=\topcolor] (0.5,0) -- ({1-\radius},0) arc (-90:0:\radius) -- (1,{1-\radius}) arc (0:90:\radius) -- (\radius,1) arc (90:180:\radius) -- (0,\radius) arc (180:270:\radius) -- cycle;%
            \draw [canvas is yx plane at z=1.5,shift={(-0.5,0.5)},fill=\topcolor] (0.5,0) -- ({1-\radius},0) arc (-90:0:\radius) -- (1,{1-\radius}) arc (0:90:\radius) -- (\radius,1) arc (90:180:\radius) -- (0,\radius) arc (180:270:\radius) -- cycle;%
        \end{scope}%
    \end{tikzpicture}%
    }%

%% file: c0_no_lab.tex
\begin{tikzpicture}

\def\squareSize{1}

\draw[thick, fill=wc] (-\squareSize, 3*\squareSize) rectangle ++(\squareSize, \squareSize);
\draw[thick, fill=gc] (-\squareSize, 2*\squareSize) rectangle ++(\squareSize, \squareSize);
\draw[thick, fill=gc] (-\squareSize, 1*\squareSize) rectangle ++(\squareSize, \squareSize);
\draw[thick, fill=wc] (-\squareSize, 0*\squareSize) rectangle ++(\squareSize, \squareSize);

\draw[thick, fill=rc] (0, 3*\squareSize) rectangle ++(\squareSize, \squareSize);
\draw[thick, fill=yc] (0, 2*\squareSize) rectangle ++(\squareSize, \squareSize);
\draw[thick, fill=yc] (0, 1*\squareSize) rectangle ++(\squareSize, \squareSize);
\draw[thick, fill=orange] (0, 0*\squareSize) rectangle ++(\squareSize, \squareSize);

\draw[thick, fill=rc] (\squareSize, 3*\squareSize) rectangle ++(\squareSize, \squareSize);
\draw[thick, fill=yc] (\squareSize, 2*\squareSize) rectangle ++(\squareSize, \squareSize);
\draw[thick, fill=yc] (\squareSize, 1*\squareSize) rectangle ++(\squareSize, \squareSize);
\draw[thick, fill=orange] (\squareSize, 0*\squareSize) rectangle ++(\squareSize, \squareSize);

\draw[thick, fill=wc] (2*\squareSize, 3*\squareSize) rectangle ++(\squareSize, \squareSize);
\draw[thick, fill=bc] (2*\squareSize, 2*\squareSize) rectangle ++(\squareSize, \squareSize);
\draw[thick, fill=bc] (2*\squareSize, 1*\squareSize) rectangle ++(\squareSize, \squareSize);
\draw[thick, fill=wc] (2*\squareSize, 0*\squareSize) rectangle ++(\squareSize, \squareSize);

\end{tikzpicture}

%% file: c0.tex
\begin{tikzpicture}

\def\squareSize{1}

\node at (-0.65,4.4*\squareSize) { \Large $\ket{0}$};
\draw[thick, fill=wc] (-\squareSize, 3*\squareSize) rectangle ++(\squareSize, \squareSize);
\draw[thick, fill=gc] (-\squareSize, 2*\squareSize) rectangle ++(\squareSize, \squareSize);
\draw[thick, fill=gc] (-\squareSize, 1*\squareSize) rectangle ++(\squareSize, \squareSize);
\draw[thick, fill=wc] (-\squareSize, 0*\squareSize) rectangle ++(\squareSize, \squareSize);

\draw[thick, fill=rc] (0, 3*\squareSize) rectangle ++(\squareSize, \squareSize);
\draw[thick, fill=yc] (0, 2*\squareSize) rectangle ++(\squareSize, \squareSize);
\draw[thick, fill=yc] (0, 1*\squareSize) rectangle ++(\squareSize, \squareSize);
\draw[thick, fill=oc] (0, 0*\squareSize) rectangle ++(\squareSize, \squareSize);

\draw[thick, fill=rc] (\squareSize, 3*\squareSize) rectangle ++(\squareSize, \squareSize);
\draw[thick, fill=yc] (\squareSize, 2*\squareSize) rectangle ++(\squareSize, \squareSize);
\draw[thick, fill=yc] (\squareSize, 1*\squareSize) rectangle ++(\squareSize, \squareSize);
\draw[thick, fill=oc] (\squareSize, 0*\squareSize) rectangle ++(\squareSize, \squareSize);

\draw[thick, fill=wc] (2*\squareSize, 3*\squareSize) rectangle ++(\squareSize, \squareSize);
\draw[thick, fill=bc] (2*\squareSize, 2*\squareSize) rectangle ++(\squareSize, \squareSize);
\draw[thick, fill=bc] (2*\squareSize, 1*\squareSize) rectangle ++(\squareSize, \squareSize);
\draw[thick, fill=wc] (2*\squareSize, 0*\squareSize) rectangle ++(\squareSize, \squareSize);

\draw[line width=1.7pt] (-\squareSize, 2*\squareSize) rectangle ++(2*\squareSize, 2*\squareSize);
\draw[line width=1.7pt] (\squareSize, 2*\squareSize) rectangle ++(2*\squareSize, 2*\squareSize);
\draw[line width=1.7pt] (-\squareSize, 0*\squareSize) rectangle ++(2*\squareSize, 2*\squareSize);
\draw[line width=1.7pt] (\squareSize, 0*\squareSize) rectangle ++(2*\squareSize, 2*\squareSize);

\end{tikzpicture}

%% file: c1.tex
\begin{tikzpicture}

\def\squareSize{1}

\node at (-0.65,4.4*\squareSize) { \Large $\ket{1}$};
\draw[thick, fill=yc] (-\squareSize, 3*\squareSize) rectangle ++(\squareSize, \squareSize);
\draw[thick, fill=bc] (-\squareSize, 2*\squareSize) rectangle ++(\squareSize, \squareSize);
\draw[thick, fill=gc] (-\squareSize, 1*\squareSize) rectangle ++(\squareSize, \squareSize);
\draw[thick, fill=wc] (-\squareSize, 0*\squareSize) rectangle ++(\squareSize, \squareSize);

\draw[thick, fill=rc] (0, 3*\squareSize) rectangle ++(\squareSize, \squareSize);
\draw[thick, fill=wc] (0, 2*\squareSize) rectangle ++(\squareSize, \squareSize);
\draw[thick, fill=yc] (0, 1*\squareSize) rectangle ++(\squareSize, \squareSize);
\draw[thick, fill=oc] (0, 0*\squareSize) rectangle ++(\squareSize, \squareSize);

\draw[thick, fill=rc] (\squareSize, 3*\squareSize) rectangle ++(\squareSize, \squareSize);
\draw[thick, fill=wc] (\squareSize, 2*\squareSize) rectangle ++(\squareSize, \squareSize);
\draw[thick, fill=yc] (\squareSize, 1*\squareSize) rectangle ++(\squareSize, \squareSize);
\draw[thick, fill=oc] (\squareSize, 0*\squareSize) rectangle ++(\squareSize, \squareSize);

\draw[thick, fill=yc] (2*\squareSize, 3*\squareSize) rectangle ++(\squareSize, \squareSize);
\draw[thick, fill=gc] (2*\squareSize, 2*\squareSize) rectangle ++(\squareSize, \squareSize);
\draw[thick, fill=bc] (2*\squareSize, 1*\squareSize) rectangle ++(\squareSize, \squareSize);
\draw[thick, fill=wc] (2*\squareSize, 0*\squareSize) rectangle ++(\squareSize, \squareSize);

\end{tikzpicture}

%% file: c2.tex
\begin{tikzpicture}

\def\squareSize{1}

\node at (-0.65,4.4*\squareSize) { \Large $\ket{2}$};
\draw[thick, fill=wc] (-\squareSize, 3*\squareSize) rectangle++(\squareSize, \squareSize);
\draw[thick, fill=gc] (-\squareSize, 2*\squareSize) rectangle++(\squareSize, \squareSize);
\draw[thick, fill=gc] (-\squareSize, 1*\squareSize) rectangle++(\squareSize, \squareSize);
\draw[thick, fill=wc] (-\squareSize, 0*\squareSize) rectangle++(\squareSize, \squareSize);

\draw[thick, fill=rc] (0, 3*\squareSize) rectangle ++(\squareSize, \squareSize);
\draw[thick, fill=yc] (0, 2*\squareSize) rectangle ++(\squareSize, \squareSize);
\draw[thick, fill=yc] (0, 1*\squareSize) rectangle ++(\squareSize, \squareSize);
\draw[thick, fill=oc] (0, 0*\squareSize) rectangle ++(\squareSize, \squareSize);

\draw[thick, fill=oc] (\squareSize, 3*\squareSize) rectangle ++(\squareSize, \squareSize);
\draw[thick, fill=wc] (\squareSize, 2*\squareSize) rectangle ++(\squareSize, \squareSize);
\draw[thick, fill=wc] (\squareSize, 1*\squareSize) rectangle ++(\squareSize, \squareSize);
\draw[thick, fill=rc] (\squareSize, 0*\squareSize) rectangle ++(\squareSize, \squareSize);

\draw[thick, fill=yc] (2*\squareSize, 3*\squareSize) rectangle ++(\squareSize, \squareSize);
\draw[thick, fill=bc] (2*\squareSize, 2*\squareSize) rectangle ++(\squareSize, \squareSize);
\draw[thick, fill=bc] (2*\squareSize, 1*\squareSize) rectangle ++(\squareSize, \squareSize);
\draw[thick, fill=yc] (2*\squareSize, 0*\squareSize) rectangle ++(\squareSize, \squareSize);

\end{tikzpicture}

%% file: c3.tex
\begin{tikzpicture}

\def\squareSize{1}

\node at (-0.65,4.4*\squareSize) { \Large $\ket{3}$};
\draw[thick, fill=yc] (-\squareSize, 3*\squareSize) rectangle ++(\squareSize, \squareSize);
\draw[thick, fill=bc] (-\squareSize, 2*\squareSize) rectangle ++(\squareSize, \squareSize);
\draw[thick, fill=gc] (-\squareSize, 1*\squareSize) rectangle ++(\squareSize, \squareSize);
\draw[thick, fill=wc] (-\squareSize, 0*\squareSize) rectangle ++(\squareSize, \squareSize);

\draw[thick, fill=rc] (0, 3*\squareSize) rectangle ++(\squareSize, \squareSize);
\draw[thick, fill=wc] (0, 2*\squareSize) rectangle ++(\squareSize, \squareSize);
\draw[thick, fill=yc] (0, 1*\squareSize) rectangle ++(\squareSize, \squareSize);
\draw[thick, fill=oc] (0, 0*\squareSize) rectangle ++(\squareSize, \squareSize);

\draw[thick, fill=oc] (\squareSize, 3*\squareSize) rectangle ++(\squareSize, \squareSize);
\draw[thick, fill=wc] (\squareSize, 2*\squareSize) rectangle ++(\squareSize, \squareSize);
\draw[thick, fill=yc] (\squareSize, 1*\squareSize) rectangle ++(\squareSize, \squareSize);
\draw[thick, fill=rc] (\squareSize, 0*\squareSize) rectangle ++(\squareSize, \squareSize);

\draw[thick, fill=yc] (2*\squareSize, 3*\squareSize) rectangle ++(\squareSize, \squareSize);
\draw[thick, fill=bc] (2*\squareSize, 2*\squareSize) rectangle ++(\squareSize, \squareSize);
\draw[thick, fill=gc] (2*\squareSize, 1*\squareSize) rectangle ++(\squareSize, \squareSize);
\draw[thick, fill=wc] (2*\squareSize, 0*\squareSize) rectangle ++(\squareSize, \squareSize);

\end{tikzpicture}

%% file: c4.tex
\begin{tikzpicture}

\def\squareSize{1}

\node at (-0.65,4.4*\squareSize) { \Large $\ket{4}$};
\draw[thick, fill=wc] (-\squareSize, 3*\squareSize) rectangle++(\squareSize, \squareSize);
\draw[thick, fill=bc] (-\squareSize, 2*\squareSize) rectangle++(\squareSize, \squareSize);
\draw[thick, fill=gc] (-\squareSize, 1*\squareSize) rectangle++(\squareSize, \squareSize);
\draw[thick, fill=wc] (-\squareSize, 0*\squareSize) rectangle++(\squareSize, \squareSize);

\draw[thick, fill=oc] (0, 3*\squareSize) rectangle ++(\squareSize, \squareSize);
\draw[thick, fill=yc] (0, 2*\squareSize) rectangle ++(\squareSize, \squareSize);
\draw[thick, fill=yc] (0, 1*\squareSize) rectangle ++(\squareSize, \squareSize);
\draw[thick, fill=oc] (0, 0*\squareSize) rectangle ++(\squareSize, \squareSize);

\draw[thick, fill=rc] (\squareSize, 3*\squareSize) rectangle ++(\squareSize, \squareSize);
\draw[thick, fill=wc] (\squareSize, 2*\squareSize) rectangle ++(\squareSize, \squareSize);
\draw[thick, fill=wc] (\squareSize, 1*\squareSize) rectangle ++(\squareSize, \squareSize);
\draw[thick, fill=rc] (\squareSize, 0*\squareSize) rectangle ++(\squareSize, \squareSize);

\draw[thick, fill=yc] (2*\squareSize, 3*\squareSize) rectangle ++(\squareSize, \squareSize);
\draw[thick, fill=gc] (2*\squareSize, 2*\squareSize) rectangle ++(\squareSize, \squareSize);
\draw[thick, fill=bc] (2*\squareSize, 1*\squareSize) rectangle ++(\squareSize, \squareSize);
\draw[thick, fill=yc] (2*\squareSize, 0*\squareSize) rectangle ++(\squareSize, \squareSize);

\end{tikzpicture}

%% file: c5.tex
\begin{tikzpicture}

\def\squareSize{1}

\node at (-0.65,4.4*\squareSize) { \Large $\ket{5}$};
\draw[thick, fill=wc] (-\squareSize, 3*\squareSize) rectangle++(\squareSize, \squareSize);
\draw[thick, fill=bc] (-\squareSize, 2*\squareSize) rectangle++(\squareSize, \squareSize);
\draw[thick, fill=gc] (-\squareSize, 1*\squareSize) rectangle++(\squareSize, \squareSize);
\draw[thick, fill=wc] (-\squareSize, 0*\squareSize) rectangle++(\squareSize, \squareSize);

\draw[thick, fill=oc] (0, 3*\squareSize) rectangle ++(\squareSize, \squareSize);
\draw[thick, fill=yc] (0, 2*\squareSize) rectangle ++(\squareSize, \squareSize);
\draw[thick, fill=yc] (0, 1*\squareSize) rectangle ++(\squareSize, \squareSize);
\draw[thick, fill=oc] (0, 0*\squareSize) rectangle ++(\squareSize, \squareSize);

\draw[thick, fill=rc] (\squareSize, 3*\squareSize) rectangle ++(\squareSize, \squareSize);
\draw[thick, fill=yc] (\squareSize, 2*\squareSize) rectangle ++(\squareSize, \squareSize);
\draw[thick, fill=yc] (\squareSize, 1*\squareSize) rectangle ++(\squareSize, \squareSize);
\draw[thick, fill=rc] (\squareSize, 0*\squareSize) rectangle ++(\squareSize, \squareSize);

\draw[thick, fill=wc] (2*\squareSize, 3*\squareSize) rectangle ++(\squareSize, \squareSize);
\draw[thick, fill=bc] (2*\squareSize, 2*\squareSize) rectangle ++(\squareSize, \squareSize);
\draw[thick, fill=gc] (2*\squareSize, 1*\squareSize) rectangle ++(\squareSize, \squareSize);
\draw[thick, fill=wc] (2*\squareSize, 0*\squareSize) rectangle ++(\squareSize, \squareSize);

\end{tikzpicture}

%% file: axis.tex
\begin{tikzpicture}
\begin{axis}[
  view={130}{20},
  axis lines=center,
  axis line style = {very thick},
  width=4cm,height=4cm,
  xtick=\empty,ytick=\empty,ztick=\empty,
  xmin=0,xmax=2,ymin=0,ymax=2,zmin=0,zmax=2,
  xlabel={\Large $x$},ylabel={\Large $y$},zlabel={\Large $z$},
  tick style={draw=none},
]
\end{axis}
\end{tikzpicture}

%% file: cube_2x2x1_labels.tex
\pgfmathsetmacro\radius{0.1}%

\def\sidecolor{bc}%
\def\frontcolor{yc}%
\def\topcolor{rc}%
\foreach \myPsi in {130}{%
    \tdplotsetmaincoords{70}{\myPsi}%
    \begin{tikzpicture}[line join=round]%
        \begin{scope}[tdplot_main_coords]%
            \filldraw [canvas is yz plane at x=1.5] (-0.5,-0.5) rectangle (1.5,1.5);%
            \filldraw [canvas is xz plane at y=1.5] (0.5,-0.5) rectangle (1.5,1.5);%
            \filldraw [canvas is yx plane at z=1.5] (-0.5,0.5) rectangle (1.5,1.5);%
            \foreach \X in {-0.5,0.5}{%
                \foreach \Y in {-0.5,0.5}{%
                   \draw [canvas is yz plane at x=1.5,shift={(\X,\Y)},fill=\frontcolor] (0.5,0) -- ({1-\radius},0) arc (-90:0:\radius) -- (1,{1-\radius}) arc (0:90:\radius) -- (\radius,1) arc (90:180:\radius) -- (0,\radius) arc (180:270:\radius) -- cycle;}}
            \draw [canvas is xz plane at y=1.5,shift={(0.5,-0.5)},fill=\sidecolor] (0.5,0) -- ({1-\radius},0) arc (-90:0:\radius) -- (1,{1-\radius}) arc (0:90:\radius) -- (\radius,1) arc (90:180:\radius) -- (0,\radius) arc (180:270:\radius) -- cycle;%
            \draw [canvas is xz plane at y=1.5,shift={(0.5,0.5)},fill=\sidecolor] (0.5,0) -- ({1-\radius},0) arc (-90:0:\radius) -- (1,{1-\radius}) arc (0:90:\radius) -- (\radius,1) arc (90:180:\radius) -- (0,\radius) arc (180:270:\radius) -- cycle;%
            \draw [canvas is yx plane at z=1.5,shift={(0.5,0.5)},fill=\topcolor] (0.5,0) -- ({1-\radius},0) arc (-90:0:\radius) -- (1,{1-\radius}) arc (0:90:\radius) -- (\radius,1) arc (90:180:\radius) -- (0,\radius) arc (180:270:\radius) -- cycle;%
            \draw [canvas is yx plane at z=1.5,shift={(-0.5,0.5)},fill=\topcolor] (0.5,0) -- ({1-\radius},0) arc (-90:0:\radius) -- (1,{1-\radius}) arc (0:90:\radius) -- (\radius,1) arc (90:180:\radius) -- (0,\radius) arc (180:270:\radius) -- cycle;%
        \end{scope}%
        \node at (-0.57,1.13) {\scriptsize$\mathbf{r}_{\noexpand\text{1}}$};
        \node at (0.18,0.93) {\scriptsize$\mathbf{r}_{\noexpand\text{2}}$};
        \node at (-1,0.5) {\scriptsize $\mathbf{r}_{\noexpand\text{3}}$};
        \node at (-0.2,0.325) {\scriptsize$\mathbf{r}_{\noexpand\text{4}}$};
        \node at (0.53,0.3) { \scriptsize $\mathbf{r}_{\noexpand\text{15}}$};
        \node at (-1,-0.4) {\scriptsize $\mathbf{r}_{\noexpand\text{5}}$};
        \node at (-0.2,-0.575) {\scriptsize $\mathbf{r}_{\noexpand\text{6}}$};
        \node at (0.53,-0.6) {\scriptsize$\mathbf{r}_{\noexpand\text{16}}$};
    \end{tikzpicture}%
    }%

%% file: b_lab.tex
\begin{tikzpicture}

\def\squareSize{1}

\draw[thick] (-\squareSize, 2*\squareSize) rectangle node {\Large $\mathbf{r}_{13}$}++(\squareSize, \squareSize);
\draw[thick] (-\squareSize, 1*\squareSize) rectangle node {\Large $\mathbf{r}_{14}$}++(\squareSize, \squareSize);

\draw[thick] (0, 5*\squareSize) rectangle node {\Large $\mathbf{r}_9$} ++(\squareSize, \squareSize) ;

\draw[thick] (0, 4*\squareSize) rectangle node {\Large $\mathbf{r}_{11}$} ++(\squareSize, \squareSize);
\draw[thick] (0, 3*\squareSize) rectangle node {\Large $\mathbf{r}_1$} ++(\squareSize, \squareSize);
\draw[thick] (0, 2*\squareSize) rectangle node {\Large $\mathbf{r}_3$} ++(\squareSize, \squareSize);
\draw[thick] (0, 1*\squareSize) rectangle node {\Large $\mathbf{r}_{5}$}++(\squareSize, \squareSize);
\draw[thick] (0, 0*\squareSize) rectangle node {\Large $\mathbf{r}_{7}$} ++(\squareSize, \squareSize);

\draw[thick] (\squareSize, 5*\squareSize) rectangle node {\Large $\mathbf{r}_{10}$} ++(\squareSize, \squareSize);
\draw[thick] (\squareSize, 4*\squareSize) rectangle node {\Large $\mathbf{r}_{12}$} ++(\squareSize, \squareSize);
\draw[thick] (\squareSize, 3*\squareSize) rectangle node {\Large $\mathbf{r}_{2}$} ++(\squareSize, \squareSize);
\draw[thick] (\squareSize, 2*\squareSize) rectangle node {\Large $\mathbf{r}_{4}$} ++(\squareSize, \squareSize);
\draw[thick] (\squareSize, 1*\squareSize) rectangle node {\Large $\mathbf{r}_{6}$} ++(\squareSize, \squareSize);
\draw[thick] (\squareSize, 0*\squareSize) rectangle node {\Large $\mathbf{r}_{8}$} ++(\squareSize, \squareSize);

\draw[thick] (2*\squareSize, 2*\squareSize) rectangle node {\Large $\mathbf{r}_{15}$}++(\squareSize, \squareSize);
\draw[thick] (2*\squareSize, 1*\squareSize) rectangle node {\Large $\mathbf{r}_{16}$} ++(\squareSize, \squareSize);

\end{tikzpicture}